\documentclass[aip,jcp,reprint]{revtex4-1}
\usepackage{graphicx}
\usepackage{dcolumn}
\usepackage{bm}
\usepackage{amsmath}
\usepackage{amssymb}
\usepackage{float}
\usepackage{epsfig}
\usepackage{color}

\usepackage{url}

\usepackage{epstopdf}

\begin{document}

\title{A Diagrammatic Kinetic Theory of Density Fluctuations in Simple Liquids in the Overdamped Limit. I. A Long Time Scale Theory for High Density}

\author{Kevin R. Pilkiewicz}\email[Electronic address: ]{pilkman@gmail.com}
\author{Hans C. Andersen}\email[Electronic address: ]{ hca@stanford.edu}\affiliation{Department of Chemistry, Stanford University, Stanford, CA 94305}
\pacs{05.20.Dd, 05.20.Jj, 05.40.-a}

\begin{abstract}

Starting with a formally exact diagrammatic kinetic theory for the equilibrium correlation functions of particle density and current fluctuations  for a monatomic liquid, we develop a theory for high density liquids whose interatomic potential has a strongly repulsive short ranged part.  We assume that interparticle collisions via this short ranged part of the potential are sufficient to randomize the velocities of the particles on a very small time scale compared with the fundamental time scale defined as the particle diameter divided by the mean thermal velocity.  When this is the case, the graphical theory suggests that both the particle current correlation functions and the memory function of the particle density correlation function evolve on two distinct time scales, the very short time scale just mentioned and another that is much longer than the fundamental time scale.  The diagrams that describe the motion on each of these time scales are identified.  When the two time scales are very different, a dramatic simplification of the diagrammatic theory at long times takes place. We identify an irreducible memory function and a more basic function, which we call the irreducible memory kernel.  This latter function evolves on the longer time scale only and determines the time dependence of the density and current correlation functions of interest at long times. In the following paper, a simple one-loop approximation for the irreducible memory kernel is used to calculate correlation functions for a Lennard-Jones fluid at high density and a variety of temperatures.

\end{abstract}

\maketitle


\section{Introduction}

Equilibrium time correlation functions of particle density and momentum density are important to the kinetic theory of liquids\cite{BoonYip,McQuarrieBook,HansenMacDonald,MazenkoBook} because their long time behavior can be related to various bulk transport coefficients and because they can be  measured in scattering experiments and easily calculated in molecular dynamics simulations, making them useful in determining the accuracy of a theory that can calculate them. 

Time correlation functions of dynamical variables  can be expressed as the solutions of formally exact integro-differential equations that typically contain a term in which the correlation function is convoluted in time with its memory function.\cite{HMori} If the memory function is known, the correlation function can be computed easily, but the memory function is generally too complicated to be evaluated exactly. Consequently, much theoretical effort has been expended throughout the last several decades to come up with tractable approximation schemes for memory functions that lead to reasonable results for their corresponding correlation functions.

Many correlation functions of interest contain the effects of physical processes that take place on different time scales. This separation of  time scales  may allow one to express the memory function as the sum of a short time component and long time component and make approximations to the two components separately.

One of the most fruitful applications of this general strategy is the mode coupling theory of G\"otze and coworkers,\cite{Gotze1,LesHouches,GotzeSjogren1,GotzeSjogren2,GotzeSjogren3,GotzeSjogren4} who were interested in constructing a long time scale theory of a two-point density correlation function. They approximated the short time part of its memory function as a function proportional to a Dirac delta function of time. They approximated the long time part of the memory function as a four-point density correlation function   that was then further approximated self consistently as a product of two  two-point density correlation functions.  

Sj\"ogren\cite{Sjogren1,Sjogren2,Sjogren3} also developed a kinetic theory that hinged on separating the memory function into a short and long time part. Using the fully renormalized kinetic theory of Mazenko\cite{FRT1,FRT2,FRT3,MazenkoYip} as a starting point, Sj\"ogren assumed that the short  time scale dynamics were dominated by binary collisions and that the long time scale dynamics consisted of recollisions and hydrodynamic backflow and used these assumptions as the basis for his separation. This had the advantage of allowing him to derive formal expressions for both the short and long time parts of his memory function, but the cost of trying to describe the dynamics on all time scales simultaneously was that these expressions were extremely cumbersome and difficult to evaluate.

Andersen\cite{DKT1,DKT2,DKT3} much later developed a formally exact diagrammatic kinetic theory for correlation functions of phase space densities, some of whose features and results were closely related to those of Mazenko's fully renormalized kinetic theory.  The diagrammatic theory had the advantage of being able to represent both correlation functions and their memory functions in a unified way in terms of diagrams that could be ascribed straightforward physical interpretations thanks to the individual elements of the diagrams being relatively simple (even though the full diagrams in general were not). Andersen and coworkers\cite{DKT4,DKT5,DKT6} then used the physical intuition this theory afforded to derive several candidate graphical representations of the short time part of the memory function of the phase space density correlation function that could be evaluated numerically and used to compute the short time behavior of several correlation functions of interest. Each of these representations in turn implied an explicit diagrammatic series for the corresponding long time part of the memory function, but these series were always too complex to evaluate. 

In this paper we develop a new method for characterizing the multiple time scale behavior of simple atomic liquids that are in what we call the `overdamped limit'. For the following qualitative discussion, let us regard the unit of time for an atomic liquid to be the ratio of the particle diameter and the mean thermal velocity and express all times in terms of that unit. Let $\nu$ be the average rate at which a particle in the fluid experiences uncorrelated binary repulsive collisions. Then $\nu^{-1}$ is approximately the relaxation time for fluctuations in a particle's momentum at equilibrium. The overdamped limit is defined by the assumption that the short ranged repulsive forces of the particles are hard enough and the density is high enough that uncorrelated binary repulsive collisions randomize the velocities of the particles on a time scale much smaller than 1. In the overdamped limit, $\nu$ is large compared with 1, and $\nu^{-1}$ is small compared with 1.

We start from the exact diagrammatic theory for the correlation function of phase space density fluctuations and focus on the diagrammatic series for a projected propagator associated with its memory function.  After a number of renormalizations, one of which makes use of a short time theory equivalent to the generalized Enskog theory for hard spheres,  we obtain a graphical formulation for the projected propagator that shows that  its magnitude and time dependence is of the form 
\[\chi_P(t)\approx (constant)\exp(-\nu t)+\nu^{-2}h(t/\nu,\nu^{-1}),\]
 where the function $h$ has only nonnegative powers of its two arguments. The two terms vary on two very distinct time scales.  The first term decays very rapidly to 0, on a time scale of $O(\nu^{-1})$.   This is the time scale for randomization of the momentum of a particle by uncorrelated binary repulsive (or hard sphere) collisions.  The second term varies on a time scale of $O(\nu)$.  This is the time scale for a particle to diffuse a distance equal to its diameter when its self diffusion coefficient is $O(\nu^{-1})$, which is the value determined by the repulsive collisions.  If $\nu$ is large and $t\gg\nu^{-1}$ , it is reasonable to approximate the behavior of the projected propagator as $\nu^{-2}h(t/\nu,0)$. In the diagrammatic theory, this corresponds to retaining only a small subset of the diagrams in the original series for the projected propagator. These diagrams are characterized topologically, and from that point on, the theory works only with the diagrams that vary on the long time scale of $O(\nu)$ and that are lowest order in powers of $\nu^{-1}$.

We note that Szamel\cite{Szamel2}  has derived a diagrammatic kinetic theory for Brownian particles undergoing diffusive motion in a solvent, using methods similar to those used in deriving the diagrammatic kinetic theory for particle systems undergoing Hamiltonian dynamics.\cite{DKT1,DKT2,DKT3}  Szamel's results and the present results have several parallels.  However, it should be noted that Szamel's method starts from the theory of Smoluchowski dynamics in a solvent whose microscopic dynamics is not explicitly discussed, whereas the present work starts from Hamiltonian dynamics for a one component particle system and shows how the longer diffusive time scale arises from a mechanical description that includes all the shorter time scales and all the degrees of freedom of the system.

In Sec.\ \ref{sec:corrfuncdiag} we state some of the results of the diagrammatic kinetic theory  of time correlation functions of the density in single particle phase space for a monatomic fluid.  In Sec.\ \ref{sec:repsrf}, we express the potential of mean force as the sum of a short ranged repulsive part and a longer ranged part, approximate the short ranged repulsive part as a hard sphere potential, and use this to express each interaction vertex in the graphical theory as a sum of two corresponding parts.   In Sec.\ \ref{sec:Hermpolyrepr}, we introduce a Hermite polynomial representation of the momentum dependence of  diagrams. Sec.\ \ref{sec:GraphicalKineticTheoryConfiguration} develops the kinetic theory in the Hermite polynomial representation, defines a `projected propagator' $\chi_P$ in diagrammatic terms, and shows that  this propagator is directly related to observable correlation functions of interest. Sec.\ \ref{sec:ProjectedPropagator} defines $\chi_P^E$, the generalized Enskog approximation for the projected propagator,  and presents a diagrammatic expression for $\chi_P$ in terms of $\chi_P^E$. Sec.\ \ref{sec:overlimiprojprop} discusses the behavior of the diagrams in $\chi_P$ in the overdamped limit and identifies the diagrams that are most important in that limit. It also presents an especially useful diagrammatic representation of $\chi_P$ in terms of an `irreducible memory kernel,' $m_{irr}$, as well as a diagrammatic series for $m_{irr}$ that describes its slowly relaxing behavior.  (Appendix \ref{sec:calculateobservables} gives the detailed relationships that allow the correlation functions of interest to be calculated from the irreducible memory kernel.) Sec.\ \ref{sec:discussion} closes with a discussion.  In a subsequent paper,\cite{paper2} a one-loop approximation for the irreducible memory kernel is formulated and evaluated for a dense Lennard-Jones fluid, and the correlation functions that follow from that result are compared with molecular dynamics computer simulation results.

\section{Correlation functions and diagrams}\label{sec:corrfuncdiag}

\subsection{Definitions}

The system of interest is a one component classical fluid of identical point particles at equilibrium.  We consider a canonical ensemble of such systems with $N$ particles in volume $V$ with temperature $T$.  The positions and momenta of the particles are ${\mathbf r}^N=\{{\mathbf r}_1,\ldots,{\mathbf r}_N\}$ and ${\mathbf p}^N=\{{\mathbf p}_1,\ldots,{\mathbf p}_N\}$.  We define a density in \textit{single particle} phase space as
\begin{equation}
f({\mathbf R},{\mathbf P},t) \equiv \sum_{i=1}^N\delta({\mathbf R}-{\mathbf r}_i(t))\delta({\mathbf P}-{\mathbf p}_i(t)),
\label{eq:density}
\end{equation}
where $\delta$ denotes a Dirac delta function.  The fluctuation of this density from its canonical ensemble average value is
\begin{equation}\delta f({\mathbf R},{\mathbf P},t) \equiv f({\mathbf R},{\mathbf P},t)-\langle f({\mathbf R},{\mathbf P},t)\rangle,
\label{eq:densityfluctuation}
\end{equation}
where the angular brackets denote an ensemble average.  This is a dynamical variable whose value fluctuates as the particles' coordinates and momenta change with time.  The time correlation function of this dynamical variable
\[C({\mathbf R},{\mathbf P},t;{\mathbf R}^\prime,{\mathbf P}^\prime,t^\prime) \equiv\left\langle\delta f({\mathbf R},{\mathbf P},t) \delta f({\mathbf R}^\prime,{\mathbf P}^\prime,t^\prime) \right\rangle\]
is a fundamental function of interest in the kinetic theory of gases and liquids.\cite{FRT1,FRT2,FRT3,MazenkoYip,Sjogren1,Sjogren2,Sjogren3,BoonYip,MazenkoBook} It is closely related to the time correlation functions of particle density and momentum density, which are related to transport properties and to the observables in coherent neutron scattering experiments. A closely related function is the corresponding self correlation function.
\[C_s({\mathbf R},{\mathbf P},t;{\mathbf R}^\prime,{\mathbf P}^\prime,t^\prime) \equiv\sum_{i=1}^N\langle\delta f_i({\mathbf R},{\mathbf P},t)\delta f_i({\mathbf R}^\prime,{\mathbf P}^\prime,t^\prime)\rangle\]
Here \[f_i({\mathbf R},{\mathbf P},t)\equiv\delta({\mathbf R}-{\mathbf r}_i(t))\delta({\mathbf P}-{\mathbf p}_i(t))\]
$C_s$ is related to the velocity autocorrelation function  and to incoherent neutron scattering.

For $t=t^\prime$ these functions are equilibrium static correlation functions.
\begin{eqnarray*}
\lefteqn{C({\mathbf R},{\mathbf P},t;{\mathbf R}^\prime,{\mathbf P}^\prime,t) =F_1({\mathbf R},{\mathbf P};{\mathbf R}^\prime,{\mathbf P}^\prime)}
\\*
&=& \rho M_M({\mathbf P})\delta({\mathbf R}-{\mathbf R}^\prime)\delta({\mathbf P}-{\mathbf P}^\prime)
\\*
&&+\rho^2M_M({\mathbf P})M_M({\mathbf P}^\prime)\left(g({\mathbf R}-{\mathbf R}^\prime)-1\right)
\\*
\vspace{2mm} \\*
\lefteqn{C_s({\mathbf R},{\mathbf P},t;{\mathbf R}^\prime,{\mathbf P}^\prime,t) =F_{s1}({\mathbf R},{\mathbf P};{\mathbf R}^\prime,{\mathbf P}^\prime)}
\\*
&=& \rho M_M({\mathbf P})\delta({\mathbf R}-{\mathbf R}^\prime)\delta({\mathbf P}-{\mathbf P}^\prime)
\end{eqnarray*}
Here $M_M$ is the Maxwell-Boltzmann distribution of momentum, and $g$ is the usual pair correlation function.

\subsection{Diagrammatic theory of time correlation functions}\label{sec:diagrammatictheory}

In previous papers,\cite{DKT1,DKT2,DKT3,DKT4,DKT5,DKT6} we have developed a diagrammatic theory of a hierarchy of time correlation functions.  The theory defines retarded propagators $\chi$ and $\chi_s$ for the correlation functions $C$ and $C_s$, respectively,\cite{Note1} and expresses the $C$ function for positive values of  $t-t^\prime$ in terms of the $t=t^\prime$ value in the following way.
\begin{eqnarray*}
\lefteqn{C({\mathbf R},{\mathbf P},t;{\mathbf R}^\prime,{\mathbf P}^\prime,t^\prime)}
\\*
 &=&\int d{\mathbf R}^{\prime\prime}d{\mathbf P}^{\prime\prime}\,
\chi({\mathbf R},{\mathbf P},t;{\mathbf R}^{\prime\prime},{\mathbf P}^{\prime\prime},t^\prime)
\\*
&&\times F_1({\mathbf R}^{\prime\prime},{\mathbf P}^{\prime\prime};{\mathbf R}^\prime,{\mathbf P}^\prime)
\\*
\lefteqn{C_{s}({\mathbf R},{\mathbf P},t;{\mathbf R}^\prime,{\mathbf P}^\prime,t^\prime)}
\\*
 &=&\int d{\mathbf R}^{\prime\prime}d{\mathbf P}^{\prime\prime}\,
\chi_{s}({\mathbf R},{\mathbf P},t;{\mathbf R}^{\prime\prime},{\mathbf P}^{\prime\prime},t^\prime)
\\*
&&\times F_{s1}({\mathbf R}^{\prime\prime},{\mathbf P}^{\prime\prime};{\mathbf R}^\prime,{\mathbf P}^\prime)
\end{eqnarray*}
The theory gives the following diagrammatic expressions for the propagators.
\vspace{1mm}

\noindent $\chi({\mathbf R},{\mathbf P},t;{\mathbf R}^\prime,{\mathbf P}^\prime,t^\prime)=$ the sum of all topologically distinct matrix diagrams with: 

\noindent $(i)$ a left root labeled $({\mathbf R},{\mathbf P},t)$ and a right root labeled $({\mathbf R}^\prime,{\mathbf P}^\prime,t^\prime)$; 

\noindent $(ii)$ free points; 

\noindent $(iii)$ $\chi^{(0)}$ bonds; 

\noindent $(iv)$ $Q_{11}^{c1}$, $Q_{11}^{c0}$,  $Q_{12}^{c1}$, $Q_{21}^{c1}$, and $Q_{22}^{c2}$ vertices; 

\noindent such that:

\noindent $(i)$ each root is attached to a bond;

\noindent $(ii)$ each free point is attached to a bond and a vertex.\,\,$\Box$
\vspace{0.5mm} 

\noindent
$\chi_{s}
({\mathbf R},{\mathbf P},t;{\mathbf R}^\prime,{\mathbf P}^\prime,t^\prime)=
$ 
the sum of all diagrams in the series just above that have a particle path from the left root to the right root.\,\,$\Box$
\vspace{1mm} 

Fig.\ 1\begin{figure}[t]
\includegraphics[width=8.5cm,keepaspectratio=true]{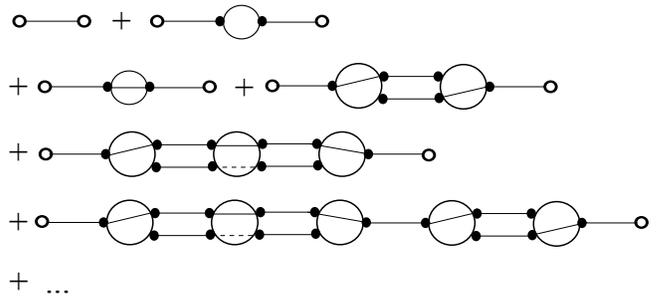}
\caption{\label{fig:chi}Some of the simplest diagrams in the series for $\chi({\mathbf R},{\mathbf P},t;{\mathbf R}^\prime,{\mathbf P}^\prime,t^\prime)$. Each root point is a small open circle. Each free point is a small closed circle. Each $\chi^{(0)}$ bond is a line. Each $Q_{nm}$ vertex is a large circle with $n$ points on the left and $m$ points on the right.  Each vertex has zero, one, or two internal lines connecting left and right points on the same vertex.  A $Q_{11}^{c1}$ vertex has an internal line, which distinguishes it from a $Q_{11}^{c0}$, which does not.  A $Q_{22}^{c2}$ vertex has two internal lines that are not topologically equivalent, so one is drawn as a solid line and the other as a dashed line. (In the actual series, each left root is labeled $({\mathbf R},{\mathbf P},t)$, and each right root is labeled $({\mathbf R}^\prime,{\mathbf P}^\prime,t^\prime)$, but these labels have been deleted from the figure for simplicity.)}
\end{figure} illustrates some of the diagrams in the series for $\chi$.
The graphs\cite{NOTE2} in this series are similar  to Mayer cluster diagrams for static correlation functions of classical fluids at equilibrium,\cite{Mayer1,MayerBook,McQuarrieBook, HansenMacDonald,MoritaHiroike,GStell,HCA1}  but they differ in several respects.  In particular, the left points and right points of a vertex or bond are not equivalent, and, to represent this, a graph should be drawn so that it is clear which points are on the left and which are on the right on every vertex or bond. (If necessary, each bond can be drawn with an arrowhead pointing from its right point to its left point.  In the figures in this paper, we adhere to the convention that the left point of a bond is always to the left of the right point, so arrowheads are not needed to make the distinction between the two nonequivalent ends of a bond.) When a free point is attached to a bond and a vertex, it is attached to the left point of the bond and a right point on the vertex, or to the right point of the bond and a left point on the vertex. When a vertex has a left  point that is also attached to a bond, we say that `the vertex has the bond on the left', with an analogous meaning to the phrase `the vertex has the bond on the right'. Some vertices have internal lines that connect right and left points on the vertex.

The $Q$ vertices describe fundamental dynamical processes, such as free particle motion and particle interactions, that affect the fluctuations of $f$.    The bonds describe the causal time evolution of fluctuations of $f$.  (Time increases from right to left in these diagrams.) Each diagram in the series above represents a single term in an iterative solution of the equations of motion for $\chi$ and $\chi_s$.
Infinite series of diagrams  such as those above can be manipulated using the topological reduction techniques developed by Morita and Hiroike.\cite{MoritaHiroike}  (See also Stell\cite{GStell} and Andersen.\cite{HCA1})

Like Mayer diagrams, each diagram has a real numerical value that is a function of the arguments assigned to the root points. Each vertex and bond in a diagram is associated with a function. To evaluate a diagram, dummy variables for position and momentum are assigned to each free point, a dummy time variable is assigned to each vertex, and an expression that contains the bond functions and vertex functions is constructed in a way that is based on the structure of the diagram.  This expression is integrated over all values of the dummy variables to give the value of the diagram.

 An overall review of the graphical kinetic theory is presented  by Ranganathan and Andersen,\cite{DKT4} including the general method for calculating the value of a diagram. A summary of much of the graphical terminology used in the theory is given by Andersen \cite{DKT1} in an appendix. Other background information is given in Appendix \ref{sec:background1}  of the present paper. The diagrammatic  theory of self functions, like $C_s$, involves the topological notion of a particle path in a diagram, which is discussed in Appendix \ref{sec:background2} of the present paper. Explicit expressions for the functions associated with the vertices and bonds in these diagrams are given in Appendix \ref{sec:bvfunctions} of the present paper.    

The fundamental theory that leads to this diagrammatic statement is a formally exact theory for the dynamics of density fluctuations in single particle phase space. A density fluctuation at a point in single particle phase space corresponds physically to the presence  \textit {or absence} of a particle at that point in the space, as should be clear from Eqs.\ (\ref{eq:density}) and (\ref{eq:densityfluctuation}). The physical interpretation can be useful for deciding how to manipulate the graphical series and for devising approximations. (The physical interpretation of diagrams is discussed in more detail by Andersen and Ranganathan.\cite{DKT4})

The exact diagrammatic theory has many more $Q$ vertices than those given in the expression above.  In the present work we have retained the $Q_{11}^{c1}$ vertex, which describes the free particle motion of a density fluctuation associated with a single particle, and the $Q_{11}^{c0}$, $Q_{21}^{c1}$, $Q_{12}^{c1}$ and $Q_{22}^{c2}$ vertices, which describe the interaction of density fluctuations associated with two distinct particles.  
The latter take into account the equilibrium structure of the fluid around the fluctuations.  Each function associated with these interaction vertices contains the equilibrium pair potential of mean force or the equilibrium direct correlation function.  The vertices we neglect are similar to those retained except that they involve equilibrium correlation functions for three or more particles.  We suspect that the vertices that are neglected represent minor corrections that do not introduce new physical effects.  (The vertices that are included have enough physics to derive the linearized Boltzmann equation for fluctuations in a dilute gas and the linearized generalized Enskog theory for fluctuations of a dense liquid, as well as describe hydrodynamic behavior for small wave vectors.) The theory is fully renormalized, in the sense of Mazenko,\cite{FRT1,FRT2,FRT3,MazenkoYip} in that the vertices retained are expressed in terms of equilibrium static correlation functions with no reference to the bare interparticle interaction. The resulting theory has a rich structure that takes into account the time evolution of arbitrarily many density fluctuations. The ultimate test of the theory will come when its predictions are compared with computer simulation results.

\section{Representation of short ranged repulsive forces}\label{sec:repsrf}

The vertices $Q_{12}^{c1}$, $Q_{22}^{c2}$, and $Q_{21}^{c1}$ contain the potential of mean force of the fluid of interest and describe density fluctuations associated with two distinct particles interacting with each other via this potential. (See Appendix \ref{sec:bvfunctions}.)   If the bare potential itself is very repulsive at short distances and the density is high, the potential of mean force will also be very repulsive at short distances. We assume the bare potential satisfies this condition. It is useful to separate the potential of mean force into its short ranged repulsive part and its longer ranged part, in the same way as the bare potential is separated in the WCA theory of equilibrium liquids,\cite{WCArepulsiveforces} because of the different physical effects of these two parts on the dynamics of the fluid. (See Appendix \ref{sec:decompositionapp} for the details.)  

Then each $Q_{12}^{c1}$, $Q_{21}^{c1}$, and $Q_{22}^{c2}$ vertex can be expressed as the sum of a repulsive ($R$) part and a longer ranged ($L$) part, and $\chi$ can be expressed in terms of these new vertices, whose functions are given in Appendix \ref{sec:decompositionapp}. 
\vspace{1mm}

\noindent$\chi(\mathbf{R}$, $\mathbf{P},t;\mathbf{R}^{\prime}$, $\mathbf{P}^{\prime},t^\prime)=$ the sum of all topologically distinct matrix diagrams with:

\noindent $(i)$ a left root labeled $(\mathbf{R}$, $\mathbf{P},t)$ and a right root labeled $(\mathbf{R}^{\prime}$, $\mathbf{P}^{\prime},t^\prime)$; 

\noindent $(ii)$ free points; 

\noindent $(iii)$ $\chi^{(0)}$ bonds; 

\noindent $(iv)$ $Q_{11}^{c1}$, $Q_{11}^{c0}$,  $Q_{12}^{c1R}$, $Q_{21}^{c1R}$, $Q_{22}^{c2R}$, $Q_{12}^{c1L}$, $Q_{21}^{c1L}$, and $Q_{22}^{c2L}$ vertices; 

\noindent such that:

\noindent $(i)$ each root is attached to a bond; 

\noindent $(ii)$ each free point is attached to a bond and a vertex.\quad$\Box$
\vspace{1mm}

Each of the functions defined graphically in this paper has a self version.  At this point, the verbal description for graphs in the self part  will begin to become lengthy and detailed, despite the fact that the actual analysis is straightforward, so henceforth we will present a graphical expression for the self part of a function only if it will appear in the final results.

The $Q^R$ vertices represent interparticle interactions that are generated by the repulsive part of the potential of mean force.  It is useful to consider the effect of completed binary collisions generated by this repulsive potential.  To do this, we start by defining a new set of vertices called $M^R$, $T_{12}^R$, $T_{21}^R$, and $T_{22}^R$, that can be used to replace the $Q^R$ vertices.  Each new vertex represents a binary collision between two particles as a result of their short ranged repulsive forces.  ($M^R$ is a memory function for repulsive collisions, and the $T^R$ vertices are scattering functions for repulsive collisions.) The procedure for   constructing these vertices is discussed in Appendix \ref{sec:repulsiveapp1}.

We are interested in interatomic potential functions for which the short ranged repulsive forces are very strong and in time scales that are long compared with the time required for  the  completion of a short ranged repulsive force collision of two particles.  We want to approximate the $M^R$ and $T^R$ vertices by the corresponding vertices for hard spheres. The procedure for doing this is discussed in Appendix \ref{sec:hardsphere}. The net result is that the $M^R$ and $T^R$ vertices are replaced by $M^H$ and $T^H$ vertices, where the $H$ denotes that they are hard sphere vertices.  Expressions for the functions associated with these vertices are given in Appendix \ref{sec:hardspherelimit}. 

The diagrammatic series for $\chi$ then becomes the following.
\vspace{1mm}

\noindent $\chi(\mathbf{R}$, $\mathbf{P},t;\mathbf{R}^{\prime}$, $\mathbf{P}^{\prime},t^\prime)=$ the sum of all topologically distinct matrix diagrams with:

\noindent $(i)$ a left root labeled $(\mathbf{R}$, $\mathbf{P},t)$ and a right root labeled $(\mathbf{R}^{\prime}$, $\mathbf{P}^{\prime},t^\prime)$; 

\noindent $(ii)$ free points; 

\noindent $(iii)$ $\chi^{(0)}$ bonds; 

\noindent $(iv)$ $Q_{11}^{c1}$, $Q_{11}^{c0}$, $M^{H}$, $Q_{12}^{c1L}$, $Q_{21}^{c1L}$, $Q_{22}^{c2L}$, $T_{12}^{H}$, $T_{21}^{H}$, and $T_{22}^{H}$  vertices; 

\noindent such that: 

\noindent $(i)$ each root is attached to a bond; 

\noindent $(ii)$ each free point is attached to a bond and a vertex.\quad$\Box$
\vspace{1mm}

\noindent The vertices with $H$ in their names are instantaneous vertices that describe completed binary hard sphere collisions.

\section{Hermite polynomial representation}\label{sec:Hermpolyrepr}

 For the development of the present theory, it is worthwhile to expand the propagator in a special complete basis set of Hermite polynomial functions of a momentum variable. Such Hermite polynomial representations of kinetic theories have been used many times.\cite{Sjogren2,MazenkoWeiYip,MazenkoYip2,DKT4,DKT5}
\begin{eqnarray}
\lefteqn{\chi(\mathbf R,\mathbf P,t;\mathbf {R}^\prime,\mathbf P^\prime,t^\prime)}
\nonumber
\\*
&=&
\sum_{\lambda\lambda^\prime}
\chi(\mathbf{R},\lambda,t;\mathbf{R}^\prime,\lambda^\prime,t^\prime)
\nonumber
\\*
&&\times h_{\lambda}(\mathbf{P})
h_{\lambda^\prime}(\mathbf{P}^\prime)M_M(\mathbf{P})M_M(\mathbf{P}^\prime)
\label{eq:chiHermite}
\end{eqnarray}
The special Hermite polynomial $h_{\lambda}(\mathbf{P})$ is defined as follows.
\begin{align}
h_{\lambda}(\mathbf{P})\equiv& [2^{\lambda_x+\lambda_y+\lambda_z}\lambda_x!\lambda_y!\lambda_z!]^{-1/2} \nonumber \\
&\times H_{\lambda_x}(\mathcal{P}_x)
H_{\lambda_y}(\mathcal{P}_y)H_{\lambda_z}(\mathcal{P}_z)
\label{eq:specialHermitepolynomial}
\end{align}
The index $\lambda$ is an ordered triplet of three nonnegative integers $(\lambda_x,\lambda_y,\lambda_z)$, which we refer to as a Hermite index. Here, the momentum $\mathcal{P}$ is equal to $(2mk_BT)^{-1/2}\mathbf{P}$, and the functions $H_n(x)$, for nonnegative integers $n$, are Hermite polynomials.

We shall refer to $\chi(\mathbf{R},\lambda,t;\mathbf{R}^\prime,\lambda^\prime,t^\prime)$ as a Hermite matrix element.  Each such matrix element can be represented as an integral containing $\chi(\mathbf{R},\mathbf{P},t;\mathbf{R}^\prime,\mathbf{P}^\prime,t^\prime)$ by using the orthonormality relationship for the basis functions.
Each vertex function and bond function in the theory can be similarly expressed in terms of matrix elements.  (See Appendix \ref{sec:Hermite} for the details.)  When it is unambiguous to do so, we will abbreviate the triplet of a Hermite index by omitting the commas, so, for example, we will write $(001)$ and $(000)$ rather than $(0,0,1)$ and $(0,0,0)$. We also use the notation $\hat0\equiv(000)$, $\hat{x}\equiv (100)$, $\hat{y}\equiv (010)$, and $\hat{z}\equiv (001)$. 

In this Hermite representation, the diagrammatic expression for $\chi$ is the same as in the momentum representation at the end of Section \ref{sec:repsrf} except for the replacement of momenta by Hermite indices.  This diagrammatic expression is given just below in Sec.\ \ref{sec:C11HermiteOriginal}.

 When a diagram is evaluated, the matrix elements of the bond and vertex functions appear as factors in an expression that is summed and integrated over dummy variables assigned to the free points, in a way very similar to the evaluation process in the momentum representation.  See Appendix \ref{sec:Hermiteevaluationapp} for a detailed statement of the rules for evaluating a diagram and an example. 

For simplicity, we adopt a notation that uses the same symbol  for the Hermite matrix element of a function as for the function itself, relying on the arguments to distinguish the two functions. (For the remainder of the paper, unless momentum arguments are explicitly indicated, the Hermite matrix representation is to be understood as being used.)

All the bonds in the theory and many of the vertices and other functions of interest are represented by graphs with one left root and one right root, and so the corresponding functions have one left Hermite index and one right Hermite index.  For such functions we occasionally use a notation in which the indices are written as subscripts.    For example, a function of the form $G({\mathbf R},\lambda,t;{\mathbf R}^\prime,\lambda^\prime,t^\prime)$ could be written as $G({\mathbf R},t;{\mathbf R}^\prime,t^\prime)_{\lambda\lambda^\prime}$.  This is especially useful since many of the quantities calculated involve matrix products of these functions.

All functions of this form that appear in the theory are translationally invariant, stationary, and retarded in time (i.e.\ the function is defined for all values of its time arguments, but the function is zero if the left time argument is earlier than the right time argument). Thus spatial Fourier transforms and Laplace transforms with regard to time can be useful, especially since many relationships among the functions are convolutions in space and/or time.  We use the following definition of the Fourier transform and Laplace transform.
\begin{equation}
\hat G({\mathbf q},t)_{\lambda\lambda^\prime}\equiv\frac{1}{V}\int_Vd{\mathbf R}d{\mathbf R}^\prime\,e^{-i{\mathbf q}\cdot({\mathbf R}-{\mathbf R}^\prime)}G(\mathbf{R},t;\mathbf{R}^{\prime},0)_{\lambda\lambda^\prime}
\label{eq:Fourier}
\end{equation}
\begin{eqnarray}\tilde G(\mathbf{R},z;\mathbf{R}^{\prime})_{\lambda\lambda^\prime} &\equiv& \int_0^\infty dt\,e^{-zt}G(\mathbf{R},t;\mathbf{R}^{\prime},0)_{\lambda\lambda^\prime}
\nonumber
\\*
\hat{\tilde G}({\mathbf q},z)_{\lambda\lambda^\prime} &\equiv&\int_{0}^\infty dt\,e^{-zt}\hat G({\mathbf q},t)_{\lambda\lambda^\prime}
\nonumber
\end{eqnarray}
with an analogous notation for Fourier transforms of functions of position only and Laplace transforms of functions of time only.

\section{Graphical kinetic theory in the Hermite representation}\label{sec:GraphicalKineticTheoryConfiguration}

\subsection{The propagator}\label{sec:C11HermiteOriginal}
The diagrammatic representation of $\chi$ in the Hermite representation is the following.
\vspace{2mm}

\noindent $\chi(\mathbf{R},\lambda,t;\mathbf{R}^{\prime},\lambda^\prime,t^\prime)=$  the sum of all topologically distinct  matrix diagrams with: 

\noindent $(i)$ a left root labeled $(\mathbf{R},\lambda,t)$ and a right root labeled $(\mathbf{R}^{\prime},\lambda^\prime,t^\prime)$; 

\noindent $(ii)$ free points; 

\noindent $(iii)$ $\chi^{(0)}$ bonds; 

\noindent $(iv)$ $Q_{11}^{c1}$, $Q_{11}^{c0}$, $Q_{12}^{c1L}$, $Q_{21}^{c1L}$, $Q_{22}^{c2L}$, $M^{H}$, $T_{12}^{H}$, $T_{21}^{H}$, and $T_{22}^{H}$  vertices; 

\noindent such that: 

\noindent $(i)$ each root is attached to a bond; 

\noindent $(ii)$ each free point is attached to a bond and a vertex.\quad$\Box$

\subsubsection{Relationship of the propagator to observables}

Some of the observable quantities of interest in the kinetic theory are the coherent intermediate scattering function, the longitudinal current correlation function, the transverse current correlation function, the incoherent intermediate scattering function, and the incoherent longitudinal current correlation function, which are all functions of wave vector and time. The last of these, for zero wave vector, is proportional to the velocity autocorrelation function. For definitions of these quantities, see, for example, Boon and Yip.\cite{BoonYip}     

The time dependence of these functions is closely related to the time dependence of the Fourier transform of some of the matrix elements of $\chi$.  It is straightforward to show that
\begin{eqnarray}
\hat{\phi}_\rho(q,t) &=&\hat{\chi}(q\hat{\mathbf k},t)_{\hat0\hat0} \label{eq:phirhodef}
\\*
\hat{\phi}_{jl}(q,t) &=&\hat{\chi}(q\hat{\mathbf k},t)_{\hat z\hat z}  \label{eq:phijldef}
\\*
\hat{\phi}_{jt}(q,t) &=&\hat{\chi}(q\hat{\mathbf k},t)_{\hat x\hat x}  \label{eq:phijtdef}
\\*
\hat{\phi}_{\rho s}(q,t) &=&\hat{\chi}_s(q\hat{\mathbf k},t)_{\hat0\hat0}  \label{eq:phirhosdef}
\\*
\hat{\phi}_{jls}(q,t) &=&\hat{\chi}_s(q\hat{\mathbf k},t)_{\hat z\hat z}  \label{eq:phijlsdef}
\end{eqnarray}
where the $\phi$ functions are the five correlation functions mentioned above, respectively, when normalized to be unity at zero time,  and $\hat{\mathbf k}$ is the unit vector in the $z$ direction.

\subsubsection{Another expression for the propagator}\label{sec:propagator2}

The Hermite matrix elements of $\chi^{(0)}$ are in Eq.\ (\ref{eq:chizerodef}).
We decompose $\chi^{(0)}$ into two parts.
Let 
\begin{eqnarray*}
\lefteqn{\chi^{(0)}({\mathbf R},\lambda,t;{\mathbf R}^\prime,\lambda^\prime,t^\prime)}
\\*
& =&\chi_{\hat0}^{(0)}({\mathbf R},\lambda,t;{\mathbf R}^\prime,\lambda^\prime,t^\prime)+\chi^{(0)}_{P}({\mathbf R},\lambda,t;{\mathbf R}^\prime,\lambda^\prime,t^\prime)
\end{eqnarray*}
where 
\[\chi^{(0)}_{\hat0}({\mathbf R},\lambda,t;{\mathbf R}^\prime,\lambda^\prime,t^\prime) =\chi^{(0)}({\mathbf R},\lambda,t;{\mathbf R}^\prime,\lambda^\prime,t^\prime)\delta(\lambda,\hat0)\]
\[\chi^{(0)}_{P}({\mathbf R},\lambda,t;{\mathbf R}^\prime,\lambda^\prime,t^\prime) =\chi^{(0)}({\mathbf R},\lambda,t;{\mathbf R}^\prime,\lambda^\prime,t^\prime)[1-\delta(\lambda,\hat0)]\]
Both of these are diagonal in their  Hermite indices.  The $\chi_{\hat0}^{(0)}$ bond function has just one nonzero element, for $\lambda=\lambda^\prime=\hat0$. 

We can use this decomposition to replace the $\chi^{(0)}$ bond by the sum of its two parts, thereby giving a series with $\chi_{\hat0}^{(0)}$ and $\chi_P^{(0)}$ bonds. A $Q^c$ matrix element all of whose Hermite indices are $\hat0$ is equal to zero. (This is a property shared by all $Q^c$ vertices that can appear in this series for  $\chi$. See Appendix \ref{sec:matrixelements} for explicit statements of the matrix elements.) If a $Q^c$ vertex in a diagram has only $\chi_{\hat0}^{(0)}$ bonds attached, the diagram will then have a value of zero.    Thus it is permissible to  include the requirement that each $Q^c$ vertex is attached to at least one $\chi_P^{(0)}$ bond without changing the value of the series.  This leads to the following graphical statement.
\vspace{2mm}

\noindent $\chi(\mathbf{R},\lambda,t;\mathbf{R}^{\prime},\lambda^{\prime},t^\prime)\equiv$ the sum of all topologically distinct matrix diagrams with:

\noindent $(i)$ a left root labeled $(\mathbf{R},\lambda,t)$ and a right root labeled $(\mathbf{R}^{\prime},\lambda^{\prime},t^\prime)$; 

\noindent $(ii)$ free points; 

\noindent $(iii)$ $\chi^{(0)}_{\hat0}$ and $\chi^{(0)}_{P}$ bonds; 

\noindent $(iv)$ $Q_{11}^{c}$, $Q_{11}^{c0}$,  $Q_{12}^{c1L}$, $Q_{21}^{c1L}$, $Q_{22}^{c2L}$, $M^H$, $T_{12}^{H}$, $T_{21}^{H}$, and $T_{22}^{H}$ vertices; 

\noindent such that:

\noindent $(i)$ each root is attached to a bond;

\noindent $(ii)$ each free point is attached to a bond and a vertex.

\noindent $(iii)$ each $Q^c$  vertex is attached to at least one $\chi_P^{(0)}$ bond.\quad$\Box$

\subsubsection{The relationship between the propagator and the coherent intermediate scattering function} \label{sec:propagatorintermediate}

To calculate the coherent intermediate scattering function using Eq.\ (\ref{eq:phirhodef}), we need an expression for the $\hat0\hat0$ matrix element of $\chi$.  The series for this matrix element can be obtained by setting $\lambda=\lambda^\prime=\hat0$ in the expression for $\chi$ in Sec.\  \ref{sec:propagator2}.  

We now analyze this graphical series  as a step toward evaluating the coherent intermediate scattering function.  The method of analysis is similar to that previously used\cite{DKT2}  to analyze a graphical series to obtain a memory function and to the analysis of graphical theories of quantum many-body phenomena to extract self-energies.\cite{Dyson} 

We examine each graph and find the $\chi^{(0)}_{\hat0}$ bonds whose removal would disconnect the diagram.  Let $n$ be the number of such bonds in a diagram.  For one graph, namely the one with a $\chi^{(0)}_{\hat0}$ bond between the roots, $n=1$.  For all other diagrams, $n\ge2$, and removal of all $n$ bonds  disconnects the diagram into two disconnected roots and $n-1$ disconnected parts, the latter of which are still internally connected.  We take each such part and replace the free point on the far left with a left root and the free point on the far right with a right root.  Both roots are assigned the Hermite index $\hat0$, because in the original $\chi$ diagram the corresponding free points were attached to  $\chi_{\hat0}^{(0)}$ bonds.  The sum of all topologically different graphs that can be obtained by this procedure defines a function to be called $M({\mathbf R},t;{\mathbf R}^\prime,t^\prime)$.
 
No graph that contributes to $M$ can contain only one vertex and no bond. Such a vertex would have been attached to only two bonds in the original diagram and both bonds would be $\chi_{\hat0}^{(0)}$ bonds.  Such diagrams do not appear in the $\chi$ series because of the last requirement in the statement of the series. Therefore every diagram in the series for $M({\mathbf R},t;{\mathbf R}^\prime,t^\prime)$ must have two or more vertices and must be a member of the following series.
\vspace{2mm}

\noindent $M(\mathbf{R},t;\mathbf{R}^{\prime},t^\prime) \equiv$ the sum of all topologically distinct matrix diagrams with:

\noindent $(i)$ a left root labeled $(\mathbf{R},\hat0,t)$ and a right root labeled $(\mathbf{R}^{\prime},\hat0,t^\prime)$; 

\noindent $(ii)$ free points; 

\noindent $(iii)$ $\chi^{(0)}_{\hat0}$ and $\chi^{(0)}_{P}$ bonds; 

\noindent $(iv)$ $Q_{11}^{c1}$, $Q_{11}^{c0}$, $Q_{12}^{c1L}$, $Q_{21}^{c1L}$, $Q_{22}^{c2L}$, $M^{H}$, $T_{12}^{H}$, $T_{21}^{H}$, and $T_{22}^{H}$ vertices; 

\noindent such that: 

\noindent $(i)$ the left root is attached to a vertex and the right root is attached to a different vertex; 

\noindent $(ii)$ each free point is attached to a bond and a vertex;

\noindent $(iii)$ each $Q^c$  vertex is attached to at least one $\chi_P^{(0)}$ bond;

\noindent $(iv)$ there is no $\chi^{(0)}_{\hat0}$ bond whose removal would disconnect the roots.\quad$\Box$
\vspace{1mm}

It follows that 
\begin{eqnarray*}
\lefteqn{\chi({\mathbf R},\hat0,t;{\mathbf R}^\prime,\hat0,t^\prime)
=\chi^{(0)}_{\hat0}({\mathbf R},\hat0,t;{\mathbf R}^{\prime},\hat0,t^{\prime})}
\\*
&&+
\int d{\mathbf R}^{\prime\prime}dt^{\prime\prime}d{\mathbf R}^{\prime\prime\prime} dt^{\prime\prime\prime}\,\chi^{(0)}_{\hat0}({\mathbf R},\hat0,t;{\mathbf R}^{\prime\prime},\hat0,t^{\prime\prime})
\\*
&&\quad\times M({\mathbf R}^{\prime\prime},t^{\prime\prime};{\mathbf R}^{\prime\prime\prime},t^{\prime\prime\prime})\chi({\mathbf R}^{\prime\prime\prime},\hat0,t^{\prime\prime\prime};\mathbf{R}^{\prime},\hat0,t^\prime)
\end{eqnarray*}
(The first term on the right is the value of the diagram in the series for $\chi$ with only one bond.  All the remaining diagrams in  $\chi$ have a $\chi_{\hat0}^{(0)}$ bond on the left, followed by a member of the series for $M$, followed by a member of the series for $\chi$ itself.  Summing over all the possibilities gives the second term on the right.)
Note that all of the functions on the right that are functions of time are retarded, and this implicitly limits the ranges of time integrations.

Taking the time derivative and Fourier transform of both sides of the equation and setting ${\mathbf q}=q\hat{\mathbf k}$ gives 
\begin{equation}
\frac{\partial \hat\chi(q\hat{\mathbf k},t)_{\hat0\hat0}}{\partial t}
=\delta(t)
+
\int_0^t dt^{\prime}\,\hat M(q\hat{\mathbf k},t-t^\prime)\hat\chi(q\hat{\mathbf k},t^{\prime})_{\hat0\hat0}
\label{eq:dchi00dt}
\end{equation}
This is a memory function equation for the propagator matrix element associated with the coherent intermediate scattering function.  Thus the function $M$ can be regarded as the memory function for the coherent intermediate scattering function.

In the graphical series for $M$, the vertex that is attached to the left root cannot have two left points. Given the way the roots are labeled with $\hat 0$ Hermite indices, the only possibility for the vertex on the left is $Q_{11}^{c1}$ because the matrix elements of $Q_{11}^{c0}$, $Q_{12}^{c1L}$, and $T_{12}^{H}$ with a left Hermite index of $\hat0$ are zero. For similar reasons, the only possibility for the vertex on the right is $Q_{11}^{c1}$ or $Q_{11}^{c0}$. Thus we have
\begin{align}
&M({\mathbf R},t;{\mathbf R}^\prime,t^\prime)
=\sum_{\lambda\lambda^\prime}\int d{\mathbf R}^{\prime\prime}d{\mathbf R}^{\prime\prime\prime}\,Q_{11}^{c1}({\mathbf R},\hat0;{\mathbf R}^{\prime\prime},\lambda)
\nonumber
\\*
&\times\chi_{P}({\mathbf R}^{\prime\prime},\lambda,t;{\mathbf R}^{\prime\prime\prime},\lambda^\prime,t^\prime) 
\nonumber
\\*
&\times  \left(Q_{11}^{c1}({\mathbf R}^{\prime\prime\prime},\lambda^\prime;{\mathbf R}^{\prime},\hat 0)+Q_{11}^{c0}({\mathbf R}^{\prime\prime\prime},\lambda^\prime;{\mathbf R}^{\prime},\hat 0)\right)
\label{eq:Mexp}
\end{align}
In a diagram for $M$, between the $Q$ on the left and the $Q$ on the right is a member of a series of diagrams that defines what shall be called the projected propagator $\chi_P$.  See the next subsection for the formal definition.  This result implies that the time dependence of the memory function associated with the  intermediate scattering function can be calculated from the time dependence of the projected propagator $\chi_P$.  (In Mori's theory of memory functions,\cite{HMori} the time dependence of a memory function of a dynamical variable is given by an operator of the form $\exp(i(I-P)L)$ that acts on the vector space of all functions in classical $N$-particle phase space.  Here $L$ is the Liouville operator, $P$ is a projection operator, and $I$ is the identity operator.  There is a close analogy to the results here, which is why we use the term `projected propagator' for $\chi_P$.)

Eq.\ (\ref{eq:Mexp}) can be written more compactly as
\begin{equation}
\hat{\tilde M}({\mathbf q},z)
=\left[\hat Q_{11}^{c1}({\mathbf q})\hat{\tilde\chi}_{P}({\mathbf q},z)\left(\hat Q_{11}^{c1}({\mathbf q})+\hat Q_{11}^{c0}({\mathbf q})\right)\right]_{\hat0\hat0}
\label{eq:MitochiP}
\end{equation}
Note that within the square brackets, the multiplications are Hermite matrix multiplications.

\subsection{The projected propagator}\label{eq:projected propagator}

Eq.\ (\ref{eq:Mexp}) follows from the diagrammatic series for $M$ in Sec.\ \ref{sec:propagatorintermediate} provided we define the projected propagator $\chi_P$  in the following way.
\vspace{1mm}

\noindent$\chi_{P}(\mathbf{R},\lambda,t;\mathbf{R}^{\prime},\lambda^{\prime},t^\prime)\equiv$ the sum of all diagrams in the latest series for $\chi(\mathbf{R},\lambda,t;\mathbf{R}^{\prime},\lambda^{\prime},t^\prime)$ that have no $\chi^{(0)}_{\hat0}$ bond whose removal would disconnect the roots.\quad$\Box$
\vspace{1mm}

The projected propagator is of central importance in the present theory.  To obtain an explicit, useful statement of its diagrammatic series, we note the following.

\noindent 1.\ In every diagram included in the series for $\chi_{P}$, each root is attached to a $\chi_{P}^{(0)}$ bond rather than a $\chi_{\hat0}^{(0)}$ bond.  
    
\noindent 2.\    An $M^H$ or $T^H$ matrix element is equal to zero unless there is at least one Hermite index on the left and one on the right that is not $\hat0$.  (See Appendix \ref{sec:matrixelements}.)  It is permissible to include the requirement that, in every diagram, every $M^H$ and $T^H$ vertex has at least one $\chi_P^{(0)}$ bond attached on its left and one on its right, since this does not change the value of the series.

We thus have the following result.
\vspace{1mm}

\noindent$\chi_{P}(\mathbf{R},\lambda,t;\mathbf{R}^{\prime},\lambda^{\prime},t^\prime)=$ the sum of all topologically distinct matrix diagrams with: 

\noindent $(i)$ a left root labeled $(\mathbf{R},\lambda,t)$ and a right root labeled $(\mathbf{R}^{\prime},\lambda^{\prime},t')$; 

\noindent $(ii)$ free points; 

\noindent $(iii)$ $\chi^{(0)}_{\hat0}$ and $\chi^{(0)}_{P}$ bonds; 

\noindent $(iv)$ $Q_{11}^{c1}$, $Q_{11}^{c0}$, $Q_{12}^{c1L}$, $Q_{21}^{c1L}$, $Q_{22}^{c2L}$, $M^H$, $T_{12}^{H}$, $T_{21}^{H}$, and $T_{22}^{H}$ vertices;

\noindent such that 

\noindent $(i)$ each root is attached to a $\chi^{(0)}_{P}$ bond; 

\noindent $(ii)$ each free point is attached to a bond and a vertex;

\noindent $(iii)$ each $Q^c$  vertex is attached to at least one $\chi_P^{(0)}$ bond;

\noindent $(iv)$ there is no $\chi^{(0)}_{\hat0}$ bond whose removal disconnects the roots;

\noindent $(v)$ each $M^H$ and $T^H$ vertex has at least one $\chi_P^{(0)}$ bond attached to its left and at least one $\chi_P^{(0)}$ attached to its right.\quad$\Box$

\subsection{The propagators associated with the longitudinal and transverse current correlation functions}

To use Eqs.\ (\ref{eq:phijldef}) and (\ref{eq:phijtdef}) to calculate the longitudinal and transverse current correlation functions, we need  expressions for the  $\hat z\hat z$ and $\hat x\hat x$ matrix elements of $\chi$.   
Consider the graphical expression for $\chi({\mathbf R},\lambda,t;{\mathbf R}^\prime,\lambda,t^\prime)$, for $\lambda=\hat z$ or $\hat x$, obtained by using the equation for $\chi$ in Sec.\ \ref{sec:propagator2}.

We perform an analysis similar to that in Sec.\  \ref{sec:propagatorintermediate} for each of these two series.  We examine each graph and find the $\chi^{(0)}_{\hat0}$ bonds whose removal would disconnect the roots.  Each graph has either no such bonds or one or more such bonds.  

A diagram in the series with no such bonds is a member of the series for $\chi_P$, and so the sum of these diagrams is $\chi_P({\mathbf R},\lambda,t;{\mathbf R}^\prime,\lambda,t^\prime)$.

If a graph has one or more such bond, we locate the leftmost and rightmost bond of this type.  (They are the same bond in the case that there is only one.)  

The part of the diagram that has these bonds and what is in between them (if there are two bonds) or the bond itself (if there is only one) is clearly a member of the graphical series for $\chi$ in which the bond attached to each root is a $\chi_{\hat0}^{(0)}$ bond.  The sum of all these possibilities is clearly equal to the $\hat 0\hat 0$ element of $\chi$ itself.  

The vertex to the immediate left of the bond on the left  must be a $Q_{11}^{c1}$ or $Q_{11}^{c0}$, vertex because these are the only vertices with one right point that have a nonzero matrix element when the right index is $\hat0$.  The part of the diagram between the left root and this vertex is a member of the series for $\chi_P$.

Similarly, the vertex to the immediate right of the bond on the right of the contribution to $\chi$ must be a $Q_{11}^{c1}$, and what is between it and the right root is a member of the series for $\chi_P$.  

We have the following exact result, which is easier to represent in the Fourier-Laplace domain.
\begin{eqnarray}
\lefteqn{\hat{\tilde\chi}(q\hat{\mathbf k},z)_{\lambda\lambda}=\hat{\tilde\chi}_P(q\hat{\mathbf k},z)_{\lambda\lambda}}
\nonumber
\\*
&&+\left[\hat{\tilde\chi}_P(q\hat{\mathbf k},z)\left(\hat{ Q}_{11}^{c1}(q\hat{\mathbf k})+\hat{Q}_{11}^{c0}(q\hat{\mathbf k})\right)\right]_{\lambda\hat0}\hat{\tilde\chi}(q\hat{\mathbf k},z)_{\hat0\hat0}
\nonumber
\\*
&&\quad\times
\left[\hat{Q}_{11}^{c1}(q\hat{\mathbf k})
\hat{\tilde\chi}_P(q\hat{\mathbf k},z)\right]_{\hat0\lambda}
\quad\text{for $\lambda=\hat z$ or $\hat x$}
\label{eq:currentpropagators}
\end{eqnarray}

\subsection{Summary of this section}

We stated the graphical expression for the propagator $\chi$ for density fluctuations in single particle phase space that incorporated the changes made in Secs.\ \ref{sec:repsrf} and \ref{sec:Hermpolyrepr}.  These included approximating the short ranged repulsive parts of the $Q$ vertices using a hard sphere model and representing the momentum dependence of diagrams in terms of Hermite matrix elements.

We showed that the $\chi_{\hat0\hat0}$, $\chi_{\hat z\hat z}$, and $\chi_{\hat x\hat x}$ Hermite matrix elements of the propagator are directly related to the coherent intermediate scattering function, the longitudinal current correlation function, and the transverse current correlation function, respectively.  

We defined a projected propagator $\chi_P$.

We found that the memory function for the kinetic equation for $\chi_{\hat0\hat0}$  can be expressed simply in terms of $\chi_P$, and therefore $\chi_{\hat0\hat0}$ can be calculated from $\chi_P$.

We found that $\chi_{\hat z\hat z}$ and $\chi_{\hat x\hat x}$ can be expressed simply in terms of $\chi_P$ and $\chi_{\hat0\hat0}$.

It follows that a theory for the projected propagator $\chi_P$ can lead to theoretical calculations of the correlation functions mentioned.

We now move forward with a theoretical analysis of $\chi_P$.

\section{Topological analysis of the series for the projected propagator}\label{sec:ProjectedPropagator}

\subsection{The generalized Enskog projected propagator}

An important approximation for $\chi_P$, and the starting point for our analysis, is the generalized Enskog projected propagator, which can be defined in the following way.
\vspace{1mm}

\noindent $\chi_{P}^{E}(\mathbf{R},\lambda,t;\mathbf{R}^{\prime},\lambda^{\prime},t^\prime)\equiv$  the sum of all topologically distinct matrix diagrams in the series in Sec.\ \ref{eq:projected propagator} for $\chi_{P}(\mathbf{R},\lambda,t;\mathbf{R}^{\prime},\lambda^{\prime},t^\prime)$ that have no $Q^L$ or $T^H$ vertices.\,\,$\Box$
\vspace{1mm}

Every diagram in this series is an alternating sequence of bonds and vertices that have one left point and one right point. Once this is recognized, it is clear that no diagram in the series has a $\chi_{\hat0}^{(0)}$ bond, since that would violate the topological restriction $(iii)$ in the diagrammatic expression for $\chi_P$. Finally, no diagram can have a $Q_{11}^{c0}$ vertex because such a vertex must be connected to one $\chi_{\hat0}^{(0)}$ bond if its matrix element is nonzero. Thus we have the following.
\vspace{1mm}
 
\noindent$\chi_{P}^{E}(\mathbf{R},\lambda,t;\mathbf{R}^{\prime},\lambda^{\prime},t^\prime)=$ the sum of all topologically distinct matrix diagrams with 

\noindent $(i)$ a left root labeled $(\mathbf{R},\lambda,t)$ and a right root labeled $(\mathbf{R}^{\prime},\lambda^{\prime},t')$; 

\noindent $(ii)$ free points; 

\noindent $(iii)$ $\chi^{(0)}_{P}$ bonds; 

\noindent $(iv)$ $Q_{11}^{c1}$ and $M^{H}$ vertices; 

\noindent such that 

\noindent $(i)$ the left root is attached to a  bond and the right root is attached to a  bond; 

\noindent $(ii)$ each free point is attached to a bond and a vertex.\quad$\Box$
\vspace{1mm}

\noindent$\chi_{sP}^{E}(\mathbf{R},\lambda,t;\mathbf{R}^{\prime},\lambda^{\prime},t^\prime)=$ the $\chi_{P}^{E}(\mathbf{R},\lambda,t;\mathbf{R}^{\prime},\lambda^{\prime},t^\prime)$ series just above   with $M^H$ vertices replaced by $M_s^H$ vertices.\quad$\Box$
\vspace{1mm}

If $\chi_P^E$  is used as an approximation for $\chi_P$ to calculate the memory function of the intermediate scattering function in Eq.\ (\ref{eq:Mexp}), the result is equivalent to the memory function of the generalized Enskog theory,\cite{BoonYip} as can be verified by a detailed calculation. This projected propagator contains the physics of uncorrelated binary hard sphere collisions, described by the hard sphere memory function $M^H$, with a collision frequency $\nu$ that is proportional to the particle density and the radial distribution function at contact. (See Appendix \ref{sec:hsapmfapp}.) Some of the properties of $\chi_P^E$ are discussed below in Sec.\ \ref{sec:chiPElargenu}.

\subsection{Topological reduction of the series for $\chi_{P}$}\label{sec:chiEQLTH}
The series for $\chi_{P}^{E}$ consists of simple chain diagrams containing $n(\ge1)$ $\chi^{(0)}_{P}$ bonds separated by $n-1$  $Q_{11}^{c1}$ and $M^{H}$ vertices. 
It is straightforward to perform a topological reduction of the series for $\chi_{P}$ that eliminates these chains and replaces them by $\chi_{P}^{E}$ bonds.  This topological reduction eliminates all $\chi_P^{(0)}$ bonds and $M^{H}$ vertices, as well as all  $Q_{11}^{c1}$ vertices that are not attached to at least one $\chi^{(0)}_{\hat0}$ bond.
Thus we have the following result.
\vspace{1mm}

\noindent$\chi_{P}(\mathbf{R},\lambda,t;\mathbf{R}^{\prime},\lambda^\prime,t^\prime) =$ the sum of all topologically distinct matrix diagrams with:

\noindent $(i)$ a left root labeled $(\mathbf{R},\lambda,t)$ and a right root labeled $(\mathbf{R}^{\prime},\lambda^\prime,t^\prime)$; 

\noindent $(ii)$ free points; 

\noindent $(iii)$ $\chi^{(0)}_{\hat0}$ and  $\chi^{E}_{P}$ bonds; 

\noindent $(iv)$ $Q_{11}^{c1}$,  $Q_{11}^{c0}$, $Q_{12}^{c1L}$, $Q_{21}^{c1L}$, $Q_{22}^{c2L}$, $T_{12}^{H}$, $T_{21}^{H}$, and $T_{22}^{H}$ vertices; 

\noindent such that:

\noindent $(i)$ each root is attached to a $\chi^{E}_{P}$ bond; 

\noindent $(ii)$ each free point is attached to a bond and a vertex;

\noindent $(iii)$ there is no $\chi_{\hat 0}^{(0)}$ bond whose removal would disconnect the roots.

\noindent $(iv)$ each $Q^c$  vertex is attached to at least one $\chi_{P}^{E}$ bond;

\noindent $(v)$ each $T^H$ vertex has at least one $\chi_P^{E}$ bond attached to its left and one $\chi_P^E$ bond attached to its right;

\noindent $(vi)$ each $Q_{11}^{c1}$ vertex is attached to a $\chi_{\hat0}^{(0)}$ bond.\quad$\Box$
\vspace{1mm}

\begin{figure}[t]\includegraphics[width=8.5cm,keepaspectratio=true]{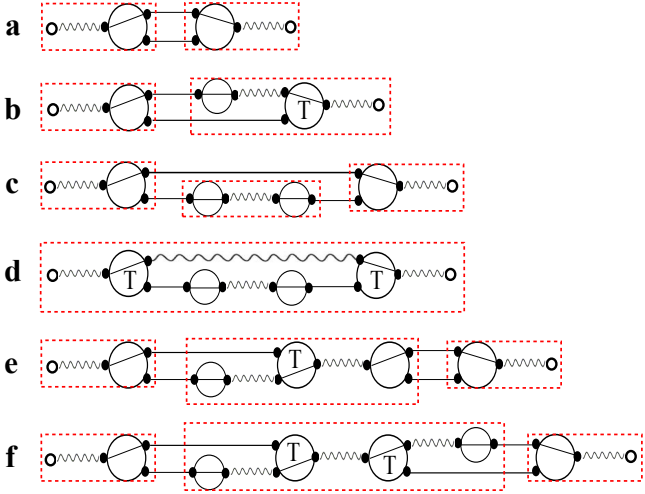}
\caption{\label{fig:chip}Several graphs in the series of Sec.\ \ref{sec:chiEQLTH} for $\chi_P$. Each small open circle is a root.  Each small closed circle is a free point.  Each large open circle is a vertex. The letter T indicates a $T^H$ vertex.  A vertex without such a letter is a $Q_{11}^c$ or $Q^{cL}$ vertex.  Each solid line between vertices is a $\chi_{\hat0}^{(0)}$ bond.  Each wavy line is a $\chi_P^E$ bond.  Some vertices have internal lines.  (Each maximal qs subgraph is surrounded by a rectangle of red dashed lines, which is not part of the diagram. Maximal qs subgraphs are discussed in Sec.\ \ref{sec:qs}.) }
\end{figure}

Fig. \ref{fig:chip} shows examples of the graphs in this series.  This will be used as the starting point for the discussion of the overdamped limit in the next section.

\section{The overdamped limit of the projected propagator}\label{sec:overlimiprojprop}

\subsection{$M^H$, $T_{21}^H$, and the hard sphere collision frequency $\nu$}\label{sec:MHproperties}

The hard sphere memory function $M^H$ that appears in the present theory  has appeared in previous kinetic theories of hard spheres, both in the context of the Boltzmann kinetic equation for the dilute gas phase and in the generalized Enskog theory for the dense liquid phase.\cite{MazenkoWeiYip,MazenkoYip2,BoonYip} It describes the effect of one binary hard sphere collision on the momentum of  a fluctuation.  Appendix \ref{sec:hardspherelimit} gives a formula for the $M^H$ function in the momentum representation, and Appendix \ref{sec:hardsphereHermite} gives an integral representation of the Hermite matrix elements of the function.  All of the results for the hard sphere fluid obtained from the present theory appear to be equivalent to those obtained by Furtado {\textit et al.}\cite{MazenkoYip2,noteFurtado}

Any matrix element of $M^H$ or $M_s^H$ with its left and/or right Hermite index equal to $\hat0$ is zero. An explicit formula for the $\hat z\hat z$ element of $M_s^H$, given in Appendix \ref{sec:hardspherelimit}, reduces to the following.
\[
\hat M_s^H
({\mathbf q})_{\hat z\hat z}
 = -\nu\]
The quantity $\nu$ has units of (time)$^{-1}$ and is positive and independent of wave vector. Except for a numerical factor of approximately unity, it is equal to the hard sphere collision frequency according to the generalized Enskog theory. Also, except for a similar numerical factor, it is equal to the reciprocal of the relaxation time associated with the decay of the velocity autocorrelation function according to that theory.  For simplicity, we shall refer to it as the generalized Enskog collision frequency.

In units in which the hard sphere diameter is approximately 1 and the mean thermal velocity is approximately 1, $\nu$ is a frequency large compared with 1, and $\nu^{-1}$ is a time small compared with 1 for atomic fluids at liquid densities.  This is because the mean free path between hard sphere collisions of a single particle is much smaller than the hard sphere diameter for liquid densities.  

  The frequency $\nu$ enters the diagrammatic theory via the $Q_{21}^{c1R}$ vertex, which describes the rate at which a propagating density fluctuation collides with an equilibrium distribution of particles.  This vertex contains a factor of $\rho g_m$ (see Eq.\ (\ref{eq:Q21R})).  In the limit that the repulsive potential becomes a hard sphere potential, $g_m$ becomes  the pair correlation function at contact for the hard spheres. (See Appendix \ref{sec:hsapmfapp}.)  This factor is the essential ingredient in  $\nu$.   Both $T_{21}^H$ and $M^H$  contain one $Q_{21}^{c1R}$ vertex and hence are $O(\nu)$.  All other hard sphere vertices are independent of the density and the pair correlation function and contain only essentially geometric information about the collisions of two hard spheres.  
 
The diagrams in the series for $\chi_P$ in Sec.\ \ref{sec:chiEQLTH} contain $T_{21}^H$ vertices but no $M^H$ vertices.  The latter have been absorbed into the $\chi_P^E$ bonds that also appear in the series.  The $T_{21}^H$ vertices generate positive powers of $\nu$ in the value of a diagram.  As we shall see below, the $\chi_P^E$ bonds lead to \textit {negative} powers of $\nu$. In the rest of this section, we explore the consequences of this and identify those diagrams that are most important for determining the behavior of $\chi_P$ for long times when $\nu$ is large.

\subsection{Properties of $\chi_P^E$}\label{sec:chiPElargenu}

The $\chi_P^E$ function represents the effect of a sequence of zero of more uncorrelated binary collisions on a density fluctuation, each of which is described by the $M^H$ memory function. 
 A matrix element of $\chi_P^E$ is nonzero only if neither of its Hermite indices is $\hat0$.
The most general diagram in $\chi_P^E$ is a chain of alternating $\chi_P^{(0)}$ bonds and vertices, and each vertex is either an $M^H$ or a $Q_{11}^{c1}$ vertex.  

$\chi_{P}^{E}({\mathbf R},\lambda,t;{\mathbf R}^\prime,\lambda^\prime,t^\prime)$ is zero for $t-t^\prime<0$ and, for very large $\nu$, decays rapidly to zero approximately as $\exp(-\nu (t-t^\prime))$ for $t-t^\prime>0$. The $\chi_P^E$ matrix is essentially an exponential of the $M^H$ matrix and represents dissipative decay of particle momentum. Every Hermite matrix element of $M^H$ contains a factor of $\nu$.

\subsection{The evaluation of diagrams for large $\nu$}

\subsubsection{Time integrations in the evaluation of a diagram}

In the series for $\chi_P$ in Sec.\ \ref{sec:chiEQLTH}, there is one diagram with no vertex, namely the diagram with a $\chi_{P}^{E}$ bond between the roots, whose value is simply $\chi_{P}^{E}(\mathbf{R},\lambda,t;\mathbf{R}',\lambda ',t')$. When evaluating every other diagram in the series, each vertex is assigned a time argument that is integrated from the initial time $t'$ to the final time $t$. (Without loss of generality, we will set $t'=0$.) The times being integrated over will appear in the integrand as the arguments of the $\chi_{\hat{0}}^{(0)}$ and $\chi_{P}^{E}$ bond functions that connect the corresponding vertices, and integrating over these bond functions will contribute factors of $t$ and $\nu^{-1}$ to the value of the diagram.

The  time dependence of the $\chi_{\hat0}^{(0)}(t;t^\prime)$ bond function  consists of a Heaviside function of its time arguments.
\begin{equation}
\chi_{\hat0}^{(0)}(t;t^\prime) =\Theta(t-t^\prime)
\label{eq:chihat0}
\end{equation}
(We have omitted the dependence on other arguments for simplicity.)  If the only bonds in a diagram were $\chi_{\hat0}^{(0)}$, then the value of the diagram would contain a factor of $t^v$, where $v$ is the number of vertices (and hence the number of time integration variables) in the diagram.  

The $\chi_P^E(t;t^\prime)$ bond function also contains a factor of $\Theta(t-t^\prime)$.  For $t-t^\prime>0$, it decreases rapidly as $t-t^\prime$ increases. Its  initial value is unity. Its approximate time dependence is an exponential decay.
\[\chi_P^E(t,t^\prime) \approx  \Theta(t-t^\prime)\exp(-\nu(t-t^\prime))\]
For the purpose of estimating the magnitude of integrals that arise from diagrams, it is helpful to replace the time dependence of this function with
\begin{equation}
\chi_P^E(t,t^\prime) \approx  \Theta\left(\nu^{-1}-(t-t^\prime)\right)\Theta(t-t^\prime)
\label{eq:chiPE}
\end{equation}
Eqs.\ (\ref{eq:chihat0}) and (\ref{eq:chiPE}) for each of the bonds in a diagram represent all the time dependent factors  that  can make the integrand zero. When these factors are nonzero, the remaining factors are bounded. The combined effect of all these Heaviside functions is to limit, in some cases severely, the ranges of values of the time variables for which the integrand is nonzero. This has two effects:  it reduces the number of powers of $t$ that the value of the diagram has, and it generates powers of $\nu^{-1}$.  

To understand the effects of the Heaviside functions on the value of  diagrams, it is helpful to introduce the idea of  quasi-simultaneous objects and subgraphs.

\subsubsection{Objects and subgraphs}

\textbf{Definition.}  An object is a root or a vertex.\,\,$\Box$\quad Thus a graph is a set of objects with bonds between them.  When a graph is evaluated, each object has a time variable assigned to it.

\textbf{Definition.} 
 A subgraph is a subset of the  objects of a graph together with all the bonds that connect one member of this subset to another. Any free points attached to these vertices and bonds are also included as part of the subgraph.\,\, $\Box$
 
   According to this definition, a subgraph can be the entire diagram or it may be just a part of the diagram. A subgraph of a diagram in  a diagrammatic series for a specific function will  typically look very much like a diagram in that series, except that it may have free points that are attached to a vertex but not a bond and it may have fewer roots than diagrams in the series have.
 
In the rest of this section, we are concerned only with the graphs that appear in the series in Section \ref{sec:chiEQLTH} for $\chi_P$ and with the properties of subgraphs of that series.  All definitions of specific types of subgraphs and statements about their properties are to be understood as applicable only in this context.

\subsubsection{Quasi-simultaneous objects and subgraphs}
\label{sec:qs}

\textbf{Definition.} Two objects in a diagram, with times $t_1$ and $t_2$ respectively, are quasi-simultaneous (henceforth abbreviated as `qs') if there is an $m$, independent of $\nu$, such that the set of Heaviside functions in Eqs.\ (\ref{eq:chihat0}) and (\ref{eq:chiPE}) for each of the bond functions in the diagram implies that the integrand for the diagram is zero for $|t_1-t_2|>m\nu^{-1}$ for all $\nu>0$.\,\,$\Box$

Two elementary properties of this concept are the following:

\noindent 1.\ If two objects are directly connected by a $\chi_P^E$ bond, they are qs.

\noindent 2. If two objects are each qs with a third object, they are qs with each other.

It is clear that if two objects are connected by a series of one or more $\chi_P^E$ bonds, they are qs.  But it is possible for two objects to be qs even if they are not connected in this way.  An example will be given below.

For a discussion of quasi-simultaneity in terms of the topology of the diagrams, see Appendix \ref{sec:qsdef}.

\textbf{Definition.}\ A qs subgraph is a subgraph in which all the  vertices and roots are qs with each other.\,\, $\Box$

\textbf{Definition.}\ A maximal qs subgraph  is a qs subgraph whose members are not qs with any object not in the subgraph.\,\, $\Box$

It is straightforward to show that:

\noindent 1.\ every object in a diagram is in one and only one maximal qs subgraph that contains at least two objects;

\noindent 2.\  every diagram in the series has a unique set of maximal qs subgraphs;

\noindent 3.\ both roots of a diagram are in the same maximal qs subgraph if and only if the diagram has only one maximal qs subgraph.

These properties are illustrated in Fig.\ \ref{fig:chip}, which has red rectangles enclosing each of the maximal qs subgraphs.  Note that within most of the rectangles, the contents are connected by wavy lines.  Not all maximal qs subgraphs are connected in this way.  In diagram d, the entire graph is one qs subgraph, despite the fact that two of the vertices are not connected to the others by wavy lines.

Appendix \ref{sec:atleasttwovertices} contains the proof of the following lemma.
 
\textbf{Lemma.} Every maximal qs subgraph in a diagram\ has at least two objects that are not $T_{21}^H$ vertices.\,\,$\Box$

\subsubsection{The $t$ and $\nu$ dependence of a diagram for large $\nu$} \label{sec:tnudepdiagrams}

We now consider how the value of a diagram depends on $t$ and $\nu$ for large $\nu$.  We separately consider diagrams whose roots are in different maximal qs subgraphs and those whose roots are in the same maximal qs subgraph.

\paragraph{The roots are in different maximal qs subgraphs.} Consider a diagram  whose two roots are in different maximal qs subgraphs. The number of maximal qs subgraphs is $W+2$,  where $W\geq 0$ is the number of maximal qs subgraphs that contain no roots. Let $v_i$ be the number of vertices, $r_i$ the number of roots, and $q_i$ the number of $T_{21}^H$ vertices in subgraph $i$, where the index $i$ goes from $1$ to $W+2$. 

When the diagram is evaluated, subgraph $i$ will have $v_i+r_i$ time arguments associated with it, and we will choose one of these arguments to be a special time argument. If the subgraph contains a root, we choose the special time to be that of the root; otherwise, we choose the special time arbitrarily. 

The contribution of each subgraph to the integrand of the diagram will be zero unless each non-special time argument is within a time of ${O}(\nu^{-1})$ of the subgraph's special time argument. Consequently, each maximal subgraph will contribute $v_i+r_i-1$ factors of $\nu^{-1}$ to the value of the diagram. In addition, as discussed in section \ref{sec:MHproperties}, each $T_{21}^{H}$ is ${O}(\nu)$, so the total number of powers of $\nu^{-1}$ that a subgraph will contribute is $v_{i}^\prime+r_i-1$, where $v_{i}^{'}\equiv v_i-q_i$ is the number of vertices in subgraph $i$ that are not $T_{21}^{H}$ vertices. We note that the lemma in Sec.\ \ref{sec:qs}  implies that $v_i^\prime+r_i\ge2$ for all maximal qs subgraphs of all diagrams in the series for $\chi_P$. 

After these integrations over the non-special time arguments, the $\chi_{\hat{0}}^{(0)}$ bonds that connect the $W+2$ subgraphs to one another will become functions of the special time arguments of the subgraphs they connect, and integrating these bond functions over the $W$ special time arguments not assigned to the roots of the diagram will generate $W$ factors of $t$. Thus the number of factors of $t$ and $\nu^{-1}$ that come from a diagram that consists of $W+2$ maximal qs subgraphs is
\[
t^W\prod_{i=1}^{W+2}\nu^{-(v_{i}^{'}+r_i-1)} \\
= \nu^{-2}\left(\nu^{-1}t\right)^W\prod_{i=1}^{W+2}\nu^{-(v_{i}^{'}+r_i-2)}
\]
  Let
\begin{equation}
N\equiv\sum_{i=1}^{W+2}(v_{i}^{'}+r_i-2)
\end{equation}
Since $v_i^\prime+r_i\ge2$ for all $i$, it follows that $N\ge0$. We have
 the following lemma.

\textbf{Lemma.} The value of a diagram in $\chi_P$ whose roots are not in the same maximal qs subgraph has the following dependence on $t$ and $\nu$.
\begin{equation}
{O}\left(\nu^{-(2+N)}\left(\nu^{-1}t\right)^W\right)
\end{equation}
where $N$ and $W$ are nonnegative integers.\,\,$\Box$

\paragraph{The roots are in the same maximal qs subgraph.}Consider a diagram that has both roots in the same maximal qs subgraph. Such a diagram has only one maximal qs subgraph. All time arguments in the diagram must be within a time of ${O}(\nu^{-1})$ of each other  if the integrand is nonzero, so  the value of the diagram decreases rapidly as a function of $t$, and the time dependence of the diagram is to a good approximation proportional to a Dirac delta function. The validity of this approximation improves as $\nu$ grows larger. A similar analysis to that used in the previous case yields the following result.

\textbf{Lemma.}\ A diagram in $\chi_P$ whose roots both lie in the same maximal qs subgraph has the following  dependence on $t$ and $\nu$.
\begin{equation}
{O}\left(\nu^{-(2+v')}\right)\delta\left(\nu^{-1}t\right)
\end{equation}
where $v^\prime\ge0$ is the number of vertices in the diagram that are not $T_{21}^H$.\,\,$\Box$\quad  (In getting this result we used the fact that $\delta(\nu^{-1}t)=\nu\delta(t)$.)

\paragraph{Final result.}

 Combining the results above, we get the following theorem.

\textbf{Theorem.} The time dependence of the projected propagator has the following form.
\begin{equation}
\chi_P(t)\sim \epsilon^{2}f(\tau,\epsilon)
\end{equation}   
where $\tau=\nu^{-1}t$ is a rescaled time  and $\epsilon =\nu^{-1}$ is a small parameter. The function $f(\tau,\epsilon)$ has the following asymptotic expansion for large $\nu$.
\begin{equation}
f(\tau,\epsilon)=\delta(\tau)\sum_{m=0}^{\infty}\epsilon^m g_m(\epsilon)+
\sum_{n=0}^{\infty}\tau^n\sum_{m=0}^{\infty}\epsilon^{m}h_{nm}(\epsilon)
\label{eq:fte}
\end{equation}
Each $g_m(\epsilon)$ and $h_{nm}(\epsilon)$ is finite in the limit of $\epsilon\rightarrow 0+$.\,\,$\Box$ 
\quad In this result, the sum is over  nonnegative $m$ because of the two previous lemmas.  

In the limit of very large $\nu$ (small $\epsilon$), we can neglect all but the leading order, $m=0$, terms in the above power series. This corresponds to neglecting all diagrams in the series for $\chi_P$ that contain a maximal qs subgraph for which $v'+r-2> 0$.  The diagrams retained are those in which every maximal qs subgraph contains exactly two objects that are not $T_{21}^H$ vertices.  We have the following theorem.

\textbf{Theorem.} For large $\nu$, the diagrammatic series for $\chi_P$ is the same as in Sec.\ \ref{sec:chiEQLTH} with the following additional restriction on the diagrams: every maximal qs subgraph contains two and only two objects that are not $T_{21}^H$ vertices.\quad$\Box$

\subsection{The topological structure of diagrams to be retained for large $\nu$} \label{sec:toporet}
 
\label{eq:initialchar}

In this section we want to characterize the diagrams to be retained for large $\nu$ in a more explicit way.  To do this, we introduce an additional topological characteristic of diagrams.

\subsubsection{$\chi_P^E$-connectivity}

\textbf{Definition.} A pair of distinct objects in a graph is $\chi_P^E$-connected if there is a sequence of objects in the graph starting with the first object of the pair and ending with the second object of the pair such that each adjacent pair of objects in the sequence is connected by a $\chi_P^E$ bond in the diagram.

\textbf{Definition.}\ A $\chi_P^E$-connected subgraph is a subgraph with two or more objects, each of whose objects is $\chi_P^E$-connected to every other object in the subgraph.\,\,$\Box$

\textbf{Definition.}\ A maximal $\chi_P^E$-connected subgraph is a $\chi_P^E$-connected subgraph that is not contained in a larger $\chi_P^E$-connected subgraph.\,\,$\Box$

\textbf{Theorem.}  If two objects are $\chi_P^E$-connected, they are in the same maximal qs subgraph.\,\,$\Box$

\textbf{Theorem.}  Every object in every graph in the series for $\chi_P$ in Sec.\ \ref{sec:chiEQLTH} is in a maximal $\chi_P^E$-connected subgraph.\,\,$\Box$  \quad The proofs of these two theorems are straightforward.

\subsubsection{A theorem about maximal qs subgraphs}
\label{sec:thmaxqs}

\textbf{Theorem.} If a maximal qs subgraph of a diagram in the series for $\chi_P$ in Sec.\ \ref{sec:chiEQLTH} has two and only two objects that are not $T_{21}^H$ vertices, it has the following properties.

 \noindent 1.\quad It is a maximal $\chi_P^E$-connected subgraph.
 
 \noindent 2.\quad It has one and only one of the following objects:  
 
 \noindent  $(i)$\ a left root that has a $\chi_P^E$ bond on the right in the original diagram, 
 
 \noindent  $(ii)$\ a $Q_{11}^{c1}$ vertex that has  no $\chi_P^{E}$ bond on the left and one on the right in the original diagram, 
 
 \noindent  $(ii)$\ a $Q_{21}^{c1L}$ vertex that has  no $\chi_P^{E}$ bond on the left and one on the right in the original diagram.  
 
\noindent 3.\quad It has one and only one of the following objects: 

 \noindent $(i)$\ a right root  that has  a $\chi_P^E$ bond on the left in the original diagram, 

 \noindent $(ii)$\ a $Q_{11}^{c1}$ or $Q_{11}^{c0}$ vertex  that has  a $\chi_P^E$ bond on the left  and none on the right in the original diagram;  

 \noindent $(iii)$\ a $Q_{22}^{c2L}$ or  $Q_{12}^{c1L}$ vertex  that has  one and only one $\chi_P^E$ bond on the left and none on the right in the original diagram.

\noindent 4.\quad It has zero or more $T_{21}^H$ vertices.  Each $T_{21}^H$ vertex has one $\chi_P^{E}$ bond on the right and one and only one $\chi_P^{E}$ bond on the left in the original diagram.

\noindent 5.\quad It has one or more $\chi_P^{E}$ bonds and no $\chi_{\hat0}^{(0)}$ bonds.

\noindent 6.\quad It has no bonds or vertices other than those mentioned above.\,\,$\Box$

The theorem is straightforward to prove by elementary methods.  The proof is a set of extensions of the reasoning used in Appendix \ref{sec:atleasttwovertices} to prove the Lemma in Sec. \ref{sec:qs}.

\textbf{Definition.} An overdamped subgraph is a subgraph that has the six properties in the theorem just above.\,\,$\Box$

The use of the term `overdamped subgraph'  is motivated by the fact that subgraphs with this property play an important role in the overdamped limit, as we shall discuss below. Using this definition, the theorem implies the following corollary.

\textbf{Theorem.} A maximal qs subgraph of a graph in the series for $\chi_P$ in Sec.\ \ref{sec:chiEQLTH} has two and only two objects that are not $T_{21}^H$ vertices if and only if it is an overdamped subgraph.\,\,$\Box$

Then the final theorem of Sec.\ \ref{sec:tnudepdiagrams} implies  the following theorem.

\textbf{Theorem.}  For large $\nu$, the diagrammatic series for $\chi_P$ is the same as in Sec.\ \ref{sec:chiEQLTH} with the following additional restriction on the diagrams: every maximal qs subgraph is an overdamped subgraph.\,\,$\Box$

\subsubsection{Properties of overdamped subgraphs}
\label{sec:propovsub}
\begin{figure}[t]
\includegraphics[width=8.5cm,keepaspectratio=true]{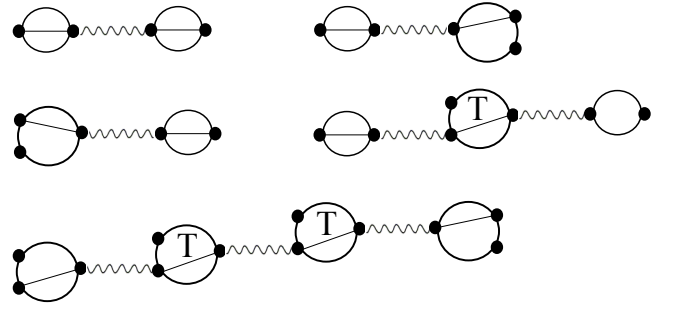}
\caption{\label{fig:odsub}Some examples of overdamped subgraphs.  See the caption of Fig.\ 2 for the meaning of the various symbols.  Each free point with no bond attached in the subgraph must be attached to a $\chi_{\hat0}^{(0)}$ bond when it appears in a graph.  These examples illustrate some general features of overdamped subgraphs that follow from the definition.  Each overdamped subgraph is a set of vertices connected by a `linear' chain of $\chi_P^E$ bonds (the wavy lines).  The left and right vertices are $Q$ vertices (large open circles with no letter inside) and the others (if any) are $T_{21}^H$ vertices (large open circles with the letter `T' inside).  There are no $\chi_{\hat0}^{(0)}$ bonds in an overdamped subgraph.  There are many other restrictions that follow from the definition.}
\end{figure}

 Some examples of overdamped subgraphs are in Fig.\ \ref{fig:odsub}.
Overdamped subgraphs have an important set of properties described in the following lemma, which follows straightforwardly from the definition above.

\textbf{Lemma.}\  The vertices in an overdamped subgraph satisfy the following restrictions, which we shall refer to as the \textit {overdamped vertex restrictions}.  These restrictions are statements about the vertices as they appear in the diagram as well as how they appear in the subgraph.

\noindent 1.\quad Every $T_{21}^H$ has two and only two $\chi_P^E$ bonds attached, one  on the left and one on the right.

\noindent 2.\quad Every $Q^{c}$ vertex has one and only one $\chi_P^E$ attached.

\noindent 3.\quad Every  $Q_{11}^{c0}$, $Q_{12}^{c1L}$, and $Q_{22}^{c2L}$ has the $\chi_P^E$ bond on the left.

\noindent 4.\quad Every  $Q_{21}^{c1L}$ has the $\chi_P^E$  bond on the right.

\noindent 5.\quad Every  $Q_{11}^{c1}$ can have the $\chi_P^E$  bond on either side.

\noindent 6.\quad No vertex is a $T_{22}^H$ or $T_{12}^H$ vertex.

\subsubsection{Characterization of diagrams to be retained for large $\nu$}\label{sec:chardiagret}

Combining the last theorem of Sec.\ \ref{sec:thmaxqs} with the lemma in Sec.\ \ref{sec:propovsub}, we obtain the following lemma.

\textbf{Lemma.}  If a diagram is in the series for $\chi_P$ for large $\nu$, every vertex in the diagram satisfies the overdamped vertex restrictions.\,\,$\Box$

From a rather complicated set of theorems, whose statements and proofs are given in Appendix \ref{sec:majorproof}, we can obtain the following lemma.

\textbf{Lemma.} If every vertex in a diagram in the series for $\chi_P$ in Sec.\ \ref{sec:chiEQLTH} satisfies the overdamped vertex restrictions, the diagram is in the series for $\chi_P$ for large $\nu$.\,\,$\Box$

Combining these two lemmas, we get the following.

\textbf{Lemma.} A diagram in the series for $\chi_P$ in Sec.\ \ref{sec:chiEQLTH} is in the series for $\chi_P$ for large $\nu$ \textit{if and only if}  every vertex in the diagram satisfies the overdamped vertex restrictions.\,\,$\Box$

We thus obtain the following characterization of the diagrams to be retained in the series for $\chi_P$ for large $\nu$.

\textbf{Theorem.}  For large $\nu$, the diagrammatic series for $\chi_P$ is the same as in Sec.\ \ref{sec:chiEQLTH} with the following additional restriction on the diagrams: every vertex satisfies the overdamped vertex restrictions.\,\,$\Box$

This is the final topological statement of the criterion for keeping diagrams in the overdamped limit.  It is stated in terms of simple topological restrictions rather than more complicated ones involving maximal qs subgraphs or maximal $\chi_P^E$-connected subgraphs.

\subsection{Overdamped theory and  overdamped limit}\label{sec:overdampedtheory}

We use the term `overdamped theory' to refer to the theory that is obtained from the diagrammatic theory when it is assumed that $\nu$ is very large and a new long time scale $\tau\equiv\nu^{-1}t$ emerges.  We use this term because a large value of $\nu$ implies that particles equilibrate their momenta much more rapidly than they can move a significant distance in configuration space.  On that time scale, a small subset of the original diagrams make the major contribution to the projected propagator.  The results for the overdamped theory are obtained by taking the `overdamped limit', which involves three actions.

1.\quad We discard the diagrams that contribute to $f(\tau,\epsilon)$ in Eq.\ (\ref{eq:fte})  to higher than zeroth order in $\epsilon$.  

2.\quad Consider the Laplace transform of the $\chi_P^E$ bond, $\tilde\chi_P^E({\mathbf R},\lambda,z;{\mathbf R}^\prime,\lambda^\prime)$.  If the Laplace transform is nonzero, it is $O(\nu^{-1})$.  It is consistent with the overdamped limit to retain only the contribution to $\tilde\chi_{P}^{E}$ that is of this order.  Thus we write
\[\tilde\chi_{P}^{E}({\mathbf R},\lambda,z;{\mathbf R}^\prime,\lambda^\prime) =\tilde\chi_{P}^{EO}({\mathbf R},\lambda,z;{\mathbf R}^\prime,\lambda^\prime)+O(\nu^{-2})\]
where the first term on the right is $O(\nu^{-1})$ for all $z$ and proportional to $\nu^{-1}$ for $z=0$. The $O$ in the superscript denotes that this is a result that is appropriate for the overdamped limit.

3.\quad We recognize that the decay of $\chi_P^E$ to zero, which occurs in a time of $O(\nu^{-1})$, is almost instantaneous.  Therefore, for simplicity, it is worthwhile replacing its bond function by a retarded Dirac delta function times a factor that gives the correct value of the time integral.  Thus we define
\[
\chi_{P}^{ED}({\mathbf R},\lambda;{\mathbf R}^\prime,\lambda^\prime)\equiv\tilde\chi_{P}^{EO}({\mathbf R},\lambda,0;{\mathbf R}^\prime,\lambda^\prime)
\]
\[
\chi_{P}^{ED}({\mathbf R},\lambda,t;{\mathbf R}^\prime,\lambda^\prime,t^\prime)\equiv\chi_{P}^{ED}({\mathbf R},\lambda;{\mathbf R}^\prime,\lambda^\prime)\delta(t-t^\prime)
\]
$\chi_{P}^{ED}({\mathbf R},\lambda,t;{\mathbf R}^\prime,\lambda^\prime,t^\prime)$ is the limiting behavior of the bond function for the $\chi_{P}^{E}({\mathbf R},\lambda,t;{\mathbf R}^\prime,\lambda^\prime,t^\prime)$ bond as the overdamped limit is approached. The $D$ in the notation is to indicate that the time dependence of this version of the function is a Dirac delta function.

The result of this process is the statement that, in the overdamped limit, $\chi_P$ approaches the following form:
\vspace{1mm}

\noindent$\chi_{P}(\mathbf{R},\lambda,t;\mathbf{R}^{\prime},\lambda^\prime,t^\prime) \to $ the sum of all topologically distinct matrix diagrams with:

\noindent $(i)$ a left root labeled $(\mathbf{R},\lambda,t)$ and a right root labeled $(\mathbf{R}^{\prime},\lambda^\prime,t^\prime)$; 

\noindent $(ii)$ free points; 

\noindent $(iii)$ $\chi^{(0)}_{\hat0}$ and  $\chi^{ED}_{P}$ bonds; 

\noindent $(iv)$ $Q_{11}^{c1}$,  $Q_{11}^{c0}$, $Q_{12}^{c1L}$, $Q_{21}^{c1L}$, $Q_{22}^{c2L}$,  and $T_{21}^{H}$ vertices; 

\noindent such that:

\noindent $(i)$ each root is attached to a $\chi^{ED}_{P}$ bond; 

\noindent $(ii)$ each free point is attached to a bond and a vertex;

\noindent $(iii)$ there is no $\chi_{\hat 0}^{(0)}$ bond whose removal would disconnect the roots;

\noindent $(iv)$\ the vertices satisfy the overdamped vertex restrictions.\quad$\Box$

This result for the overdamped limit should be compared with the series for $\chi_P$ in Sec.\ \ref{sec:chiEQLTH} just before the overdamped analysis.  The two series have many similar features, but the overdamped limit result is in many senses much simpler.  The $\chi_P^E$ bond has been replaced by the $\chi_P^{ED}$ bond, which is proportional to $\nu^{-1}$ and has a very simple Dirac delta function time dependence.  Two of the three $T^H$ vertices do not appear in the overdamped limit.  There are many more restrictions on the diagrams in the overdamped limit, and these restrictions limit the structures of the graphs to a small subset of those that appear in the series that applies before the overdamped limit. 

\subsection{Analysis of the overdamped $\chi_P$ and the irreducible memory kernel}\label{sec:chiP0dDyson}

In other theories based on physical ideas similar to those of the overdamped limit, such as the theory of colloids undergoing diffusive motion in a solvent and Markov models for dynamics in a configuration space, the memory function is related to what is called an irreducible memory function,\cite{HCAPitts,CicHess,KKawasaki,Szamel1,Szamel2} and it is worthwhile to understand and use this relationship in developing approximations for the memory function. A quantity analogous to the irreducible memory function in those theories arises straightforwardly in the present theory when a graphical analysis of the overdamped projected propagator, similar to the one performed on the propagator in Sec.\ \ref{sec:propagatorintermediate}, is performed.
In this subsection, we define an irreducible memory kernel $m_{irr}$.  In Appendix \ref{sec:calculateobservables}, $m_{irr}$ is used to define an irreducible memory function.

\begin{figure}[t]
\includegraphics[width=8.5cm,keepaspectratio=true]{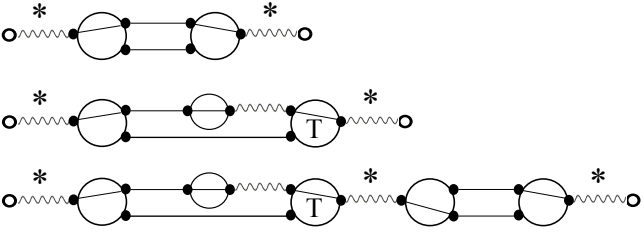}
\caption{\label{fig:chipodl}Some examples of diagrams in the series for $\chi_P$ in the overdamped limit with their nodal bonds indicated by asterisks.  See the caption of Fig.\ 2 for the meaning of the various symbols.}
\end{figure}

In the diagrammatic equation in Sec.\ \ref{sec:overdampedtheory} for the projected propagator $\chi_{P}$ in the overdamped limit, we examine each graph and find $\chi_P^{ED}$ bonds whose removal would disconnect the roots. Such a bond will be called a nodal bond.  The diagram consisting of a single $\chi_P^{ED}$ bond connecting the roots has only one such bond. Every other diagram in the series has at least two such bonds, namely the bonds attached to the roots.  Fig.\ \ref{fig:chipodl}  shows several examples of graphs in the series for $\chi_P$ in the overdamped limit with the nodal bonds indicated by asterisks.  The irreducible memory kernel will be defined as the sum of all the things that can appear `between' two such bonds.

In what appears between two adjacent  nodal bonds, the vertex on the left had a $\chi_P^{ED}$ bond (i.e.\ the nodal bond) on its left in the original diagram.  Thus it has to be either a $Q_{11}^{c1}$, $Q_{11}^{c0}$, $Q_{12}^{c1L}$, or $Q_{22}^{c2L}$ vertex.  The first, second, and fourth possibilities are not acceptable, so it must be a $Q_{12}^{c1L}$.  (To see this, note that if it were a $Q_{11}^{c1}$ or $Q_{11}^{c0}$ it would have to have a $\chi_{\hat0}^{(0)}$ on the right. Such a bond, if removed, would disconnect the roots, which is not allowed. If it were a $Q_{22}^{c2L}$, the left point would have a $\chi_{\hat0}^{(0)}$ bond on the left which provides a connection to the left root [see Appendix \ref{sec:validgraphs}], which is inconsistent with the fact that the bond that was removed was a nodal bond whose removal would disconnect the roots.)

Similarly, in what appears between two adjacent nodal bonds, the vertex on the right has a $\chi_P^{ED}$ bond on the right in the original diagram.  Therefore it must be a  $T_{21}^H$, $Q_{21}^{c1L}$, or $Q_{11}^{c1}$.  The latter possibility is unacceptable for a reason similar to that in the previous paragraph. Thus the vertex on the right must be a $T_{21}^H$ or $Q_{21}^{c1L}$ vertex. Thus we have the following.
\vspace{1mm}
 
\noindent $m_{irr}(\mathbf{R},\lambda,t;\mathbf{R}^{\prime},\lambda^\prime,t^\prime)=$ the sum of all topologically distinct matrix diagrams with:

\noindent $(i)$ a left root labeled $(\mathbf{R},\lambda,t)$ and a right root labeled $(\mathbf{R}^{\prime},\lambda^\prime,t^\prime)$; 

\noindent $(ii)$ free points; 

\noindent $(iii)$ $\chi^{(0)}_{\hat0}$ and  $\chi^{ED}_{P}$ bonds; 

\noindent $(iv)$ $Q_{11}^{c1}$,  $Q_{11}^{c0}$, $Q_{12}^{c1L}$, $Q_{21}^{c1L}$, $Q_{22}^{c2L}$,  and $T_{21}^{H}$ vertices; 

\noindent such that:

\noindent $(i)$ the left root is attached to $Q_{12}^{c1L}$ vertex; 

\noindent $(ii)$ the right root is attached to a $T_{21}^H$ or $Q_{21}^{c1L}$ vertex;

\noindent $(iii)$ each free point is attached to a bond and a vertex;

\noindent $(iv)$ there is no bond whose removal would disconnect the roots;

\noindent $(v)$\ every vertex not attached to a root satisfies the overdamped vertex restrictions;

\noindent $(vi)$ a $Q_{12}^{c1L}$ attached to the left root has no $\chi_P^{ED}$ bond on the right;

\noindent $(vii)$ a $T_{21}^H$ vertex attached to the right root has one and only one $\chi_P^{ED}$ bond on the left;

\noindent $(viii)$ a $Q_{21}^{c1L}$ vertex attached to the right root has no $\chi_P^{ED}$ bond on the left.\quad$\Box$
\vspace{1mm}

\begin{figure}[t]
\includegraphics[width=8.5cm,keepaspectratio=true]{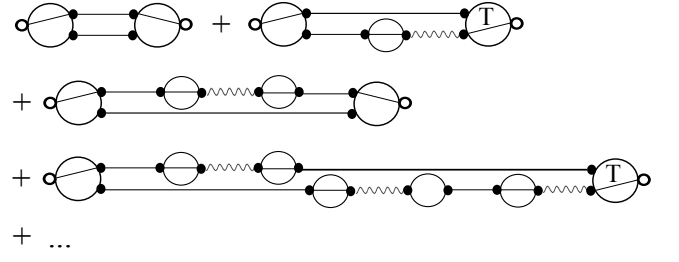}
\caption{\label{fig:mirr}  Some of the diagrams in the series for $m_{irr}$.  See the caption of Fig.\ 2 for the meaning of the various symbols.}
\end{figure}

\noindent Fig.\ \ref{fig:mirr} contains some of the diagrams in the series for $m_{irr}$.

\noindent $m_{s\,irr}(\mathbf{R},\lambda,t;\mathbf{R}^{\prime},\lambda^\prime,t^\prime)=$ the series  just above with the following changes:

\noindent $(iii)$ $\chi^{ED}_{sP}$ bonds are also allowed; 

\noindent such that:

\noindent $(v)$-$(viii)$ these restrictions apply but they apply with $\chi_P^{ED}$ replaced by $\chi_P^{ED}$ or $\chi_{sP}^{ED}$;

\noindent $(ix)$ there is a particle path from the left root to the right root;

\noindent $(x)$ a $\chi_{sP}^{ED}$ bond can appear only on the particle path just mentioned.
\,\,$\Box$
\vspace{1mm}

We obtain the following expression for the projected propagator in the overdamped limit.
\vspace{1mm}

\noindent $\chi_{P}(\mathbf{R},\lambda,t;\mathbf{R}^{\prime},\lambda^\prime,t^\prime)=$ the sum of all topologically distinct matrix diagrams with:

\noindent $(i)$ a left root labeled $(\mathbf{R},\lambda,t)$ and a right root labeled $(\mathbf{R}^{\prime},\lambda^\prime,t^\prime)$; 

\noindent $(ii)$ free points; 

\noindent $(iii)$ $\chi^{ED}_{P}$ bonds; 

\noindent $(iv)$ $m_{irr}$ vertices; 

\noindent such that:

\noindent $(i)$ the left root is attached to a  bond and the right root is attached to a  bond; 

\noindent $(ii)$ each free point is attached to a bond and a vertex.
\,\,$\Box$
\vspace{1mm}

\noindent $\chi_{sP}(\mathbf{R},\lambda,t;\mathbf{R}^{\prime},\lambda^\prime,t^\prime)=$ the series  just above with the $\chi_P^{ED}$ bonds and $m_{irr}$ vertices replaced by $\chi_{sP}^{ED}$ bonds and $m_{s\,irr}$ vertices, respectively.\,\,$\Box$
\vspace{1mm}

These diagrams are simple chains of alternating $\chi_P^{ED}$ bonds and $m_{irr}$  vertices, and each infinite series can easily be summed using Fourier-Laplace transforms.

\subsection{Comment} 

The two major results of this section are diagrammatic expressions in Sec.\ \ref{sec:chiP0dDyson} for $\chi_P$ and $\chi_{sP}$ in terms of $m_{irr}$ and $m_{s\, irr}$ in the overdamped limit and diagrammatic series for $m_{irr}$ and $m_{s\,irr}$.

The calculation of approximations for $m_{irr}$ and $m_{s\,irr}$ based on the graphical series for these functions is the major theoretical task that is required to derive an explicit form of the kinetic equations for time correlation functions of interest. The series for $m_{irr}$ and $m_{s\,irr}$ each contain infinitely many diagrams, and topological reduction techniques will be necessary for putting them into a form suitable for approximations.  This is discussed in the next paper in this series.\cite{paper2}     

\section{Discussion}\label{sec:discussion}

To use the overdamped theory to carry out calculations of the time correlation functions of a specific atomic liquid for a particular density and temperature, the following steps are required.  We shall discuss this for the case of a one component liquid, but the entire theory is generalizable to mixtures of atomic species.

1.\quad Perform equilibrium canonical ensemble simulations to obtain the radial distribution function of the fluid of interest.  

2.\quad Calculate the short ranged repulsive part of the potential of mean force from the radial distribution function, using Eqs.\ (\ref{eq:pmfdef}) and the method of Appendix \ref{sec:decompositionapp}.

3.\quad Calculate $\hat {\tilde M}_s^R({\mathbf 0},0)_{\hat z\hat z}$, the Fourier-Laplace transform of the matrix element of the self memory function for a dilute gas of particles whose \textit {bare potential} is equal to the potential constructed in step 2.  ($\hat {\tilde M}_s^R({\mathbf q},z)$ is called the RPMF approximation for the memory function of the fluid of interest.  The method for calculating this matrix element is discussed by Noah-Vanhoucke and Andersen.\cite{DKT6}    These calculations do not involve dynamical simulations of the liquid, merely trajectory calculations for collisions of two particles.)  Choose the effective hard sphere diameter $d$ associated with the repulsive part of the potential of mean force of the fluid of interest so that, at the density of the fluid of interest, the value of $\hat {\tilde M}_s^H({\mathbf 0},0)_{\hat z\hat z}$ of the hard sphere fluid is equal to $\hat {\tilde M}_s^R({\mathbf 0},0)_{\hat z\hat z}$ for the fluid of interest.  The value of $\nu$ for the fluid of interest then is equal to $-\hat {\tilde M}_s^R({\mathbf 0},0)_{\hat z\hat z}$.  (See Eq.\ \ref{eq:nu}.)

4.\quad Calculate an approximate result for the hard sphere memory functions $M^H$ and $M_s^H$ for the hard sphere fluid at the density of the fluid of interest. This is most conveniently done using the method of kinetic models.\cite{GRO59,SUG68,MazenkoWeiYip,MazenkoYip2}

5.\quad  Use the approximate hard sphere memory functions $M^H$ and $M_s^H$ to calculate the $\chi_P^{E}$ and $\chi_{sP}^{E}$ propagators of the generalized Enskog theory as well as their behavior in the overdamped limit.

6.\quad  Devise a graphical approximation for $m_{irr}$, using the graphical expression in Sec.\ \ref{sec:chiP0dDyson}.

7.\quad  Calculate numerical values, as a function of wave vector and time, of the coherent intermediate scattering function and the other observable correlation functions of interest using the method discussed in Appendix\  \ref{sec:calculateobservables}.

The theory is based on several assumptions. The first is that the short ranged repulsive part of the potential of mean force is a repulsive enough potential that its vertices can be replaced by those appropriate for a hard sphere fluid.  The second is that the quantity $\nu$, the hard sphere collision frequency, is large enough that the asymptotic theoretical result for very large $\nu$ is applicable. Whether or not these two assumptions are correct for any specific liquid depends on the interparticle potential of mean force for the liquid.  If the bare interparticle potential is `hard' enough at short distances, the first assumption will be satisfied because the repulsive part of the potential of mean force will be similar to that of the bare potential.  If the density is increased, the value of $\nu$ will increase, but it is not clear how accurate the theory will be for any particular value of $\nu$.  The third assumption is that the hard sphere kinetic model that is used is accurate enough, and the fourth is that the graphical approximation used for $m_{irr}$ is accurate enough. These assumptions depend on the kinetic  model, the graphical approximation, and the liquid under consideration.

In the accompanying paper,\cite{paper2} we present the results that follow from a simple `one-loop' graphical approximation for the irreducible memory kernel and a simple kinetic model for hard spheres. The system investigated is a dense Lennard-Jones liquid at a variety of temperatures. The results for the observable correlation functions are compared with those obtained from simulations of the Lennard-Jones liquid.

\appendix

\section{The diagrammatic theory}\label{sec:background1}

\subsection{Elementary properties of diagrams and their components}\label{sec:background2}

\textbf{Matrix diagrams.} The graphs in the present theory are `matrix' diagrams (see  Andersen \cite{DKT1}).  The term is defined by three restrictions.  

\noindent 1.\quad Each left root is attached to a left point of a vertex or bond and nothing else.  

\noindent 2.\quad Each right root is attached to the right point of a vertex or bond and nothing else.  

\noindent 3.\quad Each free point is attached to the right point of one vertex or bond and the left point of another vertex or bond and nothing else.

\textbf{Instantaneous and retarded bonds and vertices.}
All vertices in the theory are either instantaneous or retarded.
When a graph is evaluated, 
an instantaneous vertex has the same time argument assigned to all its points, and a retarded vertex has one time argument assigned to its left points and another attached to its right points.  

The vertex function for an instantaneous vertex is independent of its time argument.  The vertex function for a retarded vertex is a function of the difference of its left time and its right time, and that function is zero if the left time precedes the right time.

\textbf{Paths, forward paths, and particle paths.}
For various parts of the theory, it is worthwhile regarding a graph as a set of roots and vertices with bonds between them.  In the following, we shall use the term `object' to refer to either a root or a vertex.  Thus a graph is a set of objects with bonds between them.

A \textit {path} from object $A$ to object $B$ in a diagram is an alternating sequence of objects and bonds such that the first object in the sequence is $A$, the last object is $B$, and such that each bond in the sequence is attached to the object that precedes it and the object that follows it and no bond appears more than once in the path.

A \textit {forward path} between two objects in a diagram is a path such that  the tail end of each bond is attached to the object that precedes it in the path, and the head end of the bond is attached to the object that follows it in the path. 

A \textit {particle path} between two objects $A$ and $B$ in a diagram is a path such that:
1.\ every bond in the path is a self bond;
2.\  for every vertex in the path that is not $A$ or $B$ there is an internal line connecting the points to which the preceding bond and following bond in the path are attached. 

\subsection{Topological properties of valid graphs}\label{sec:validgraphs}
Graphical expressions like the one for $\chi$ in Sec.\ \ref{sec:diagrammatictheory} are obtained from  the diagrammatic kinetic theory.\cite{DKT3}  Each diagram obtained from this theory and every diagram obtained by subsequent analysis of the type performed in this work satisfies the following requirements\cite{ftn:noncausal}.

\noindent 1.\quad For any object in the diagram other than a left root, there is at least one forward path from that object  to a left root.

\noindent 2.\quad For any object in the diagram other than a right root, there is at least one forward path from the right root to that object.

\noindent 3.\quad There is no forward path from an object to itself.

\subsection{Differences from previous uses of the diagrammatic kinetic theory}

\textbf{Absence of symmetry numbers.}  The derivation of the form of the present theory required use of graphs that have nontrivial symmetry numbers because of symmetry properties of the vertices.  We have formulated the starting point for the theory of the overdamped limit using vertices that have no such symmetry, and the symmetry number of every diagram is 1.  Only two symmetric vertices were introduced later in the development, namely the $T_{22}^R$ and $T_{22}^H$ vertices.  Because of the other properties of graphs in which these vertices appear, the points in this vertex do not have the symmetry of the vertex itself, so the overall symmetry number of every graph is one. This simplifies the evaluation of the diagrams.

\textbf{The form of $Q_{12}^{c1}$.} The prescription for deciding which vertices to retain leads to a different form of the $Q_{12}^{c1}$ vertex from what we used in previous papers.  The result used here is in Eq.\ (\ref{eq:Q12}).  (For further discussion, see Appendix B of Andersen.\cite{DKT4})

\textbf{Notation.}  The notation in this paper is consistent with that in previous papers but in some cases it has been simplified.

\subsection{Bond and vertex functions of the original graphical formulation}\label{sec:bvfunctions}

Here and in some of the following sections, we use a notation in which, for example, `1' is used as an abbreviation for $({\mathbf R}_1,{\mathbf P}_1)$.  Thus the first equation below is an abbreviation for the following equation. 
\[\chi^{(0)}({\mathbf R}_1,{\mathbf P}_1,t;{\mathbf R}_1^\prime,{\mathbf P}_1^\prime, t^\prime) =\Theta(t-t^\prime)\delta({\mathbf R}_1-{\mathbf R}_1^\prime)\delta({\mathbf P}_1-{\mathbf P}_1^\prime)\]
$\Theta$ denotes the Heaviside function. Some functions depend on only some of their arguments.   For example, $g(12)$ is the same as $g({\mathbf R}_1,{\mathbf R}_2)$ (often written as $g(r)$ where $r=|{\mathbf R}_1-{\mathbf R}_2|$), and $M_M(1)$ is the same as $M_M({\mathbf P}_1)$.  $\nabla_R$ denotes the gradient with regard to the \textit {first} position argument of a function, and $\nabla_P$ denotes the gradient with regard to the \textit {first} momentum argument.
If a function has more than one argument abbreviated as an integer, commas will not be used to separate the arguments. In formulas for vertex functions and bond functions, however, the left arguments are separated from the right arguments by a semicolon.

\textbf{Unperturbed propagator}
\[\chi^{(0)}(1,t;1^\prime, t^\prime) =\Theta(t-t^\prime)\delta(11^\prime)\]

\textbf{$Q$ vertices}
\begin{eqnarray}
Q_{11}^{c1}(1;1^\prime) &=&-({\mathbf P}_1/m)\cdot\nabla_R\delta(11^\prime)
\\
Q_{11}^{c0}(1;1^\prime) &=&\rho M_M(1)({\mathbf P}_1/m)\cdot\nabla_Rc(11^\prime)
\\
Q_{12}^{c1}(1;1^\prime2^\prime)
 &=&\nabla_Rv(1^\prime2^\prime)\cdot\nabla_P\delta(11^\prime)
\label{eq:Q12}
\\
Q_{21}^{c1}(12;1^\prime) &=&\rho g(12) \left[{M_M(1)M_M(2)/ M_M(1^\prime)}\right]
\nonumber
\\*
&&\times\nabla_Rv(12)\cdot\nabla_P\delta(11^\prime)
\label{eq:Q21}
\\
Q_{22}^{c2}(12;1^\prime2^\prime) &=&\nabla_Rv(1^\prime2^\prime)\cdot\nabla_P\delta(11^\prime)\delta(22^\prime)
\label{eq:Q22}
\end{eqnarray} 
Here $v(12)$ is the potential of mean force, defined as  
\begin{equation}
v(r) = -k_BT\ln g(r),
\label{eq:pmfdef}
\end{equation}
and  $c(11^\prime)$ is the direct correlation function.

$Q_{11}^{c1}$, $Q_{12}^{c1}$, and $Q_{21}^{c1}$ each have a single internal line connecting the first left point and the first right point.  $Q_{22}^{c2}$ has two internal lines, one connecting the first left point and the first right point, and another connecting the second left point and the second right point.  None of these vertices has any symmetry.  For $Q_{22}^{c2}$, the two internal lines are not equivalent, so the two lines should be drawn in different ways.  For example, the first line could be a solid line and the second could be a dotted line.

\section{The representation of repulsive forces}

\subsection{Decomposition of the potential of mean force}\label{sec:decompositionapp}

Let $r_m$ and $v_m$ be the distance of the first minimum of $v$ (the potential of mean force) and the value of $v$ at the minimum, respectively.  Let $g_m$ be the value of the pair correlation function $g(r)$ at that distance, which corresponds to the location of the first maximum of $g$. 
We separate $v$ using the WCA prescription\cite{WCArepulsiveforces} in the following way.  
\[v(12) =v^R(12)+v^L(12)\]
The superscripts $R$ and $L$ refer to
 the short ranged repulsive part and  the longer ranged part, respectively.  (For dense liquids, the longer ranged part is oscillatory and generates both attractive and repulsive forces.)  Here
 \begin{align*}
 v^R(12) &= v(12) -v_m&&\text{for $R_{12}\le r_m$}
 \\*
 &=0&&\text{for $R_{12}\ge r_m$}
 \\*
 v^L(12) &= v_m&&\text{for $R_{12}\le r_m$}
 \\*
 &=v(12)&&\text{for $R_{12}\ge r_m$}
  \end{align*}  
We define
\begin{align*}
g^R(12) &\equiv\exp(-v^R(12)/k_BT)
\\*
g^L(12) &\equiv\exp(-v^L(12)/k_BT)
\end{align*}
and hence
\[g(12) =g^R(12)g^L(12)\]

The separation of the potential of mean force into two parts leads to the separation of  $Q_{12}^{c1}$, $Q_{22}^{c2}$, and $Q_{21}^{c1}$ into short ranged repulsive and longer ranged parts.
\begin{eqnarray}
Q_{12}^{c1}(1;1^\prime2^\prime) &=&Q_{12}^{c1R}(1;1^\prime2^\prime)+Q_{12}^{c1L}(1;1^\prime2^\prime)
\label{eq:decompQ12}
\\*
Q_{21}^{c1}(12;1^\prime) &=&Q_{21}^{c1R}(12;1^\prime)+Q_{21}^{c1L}(12;1^\prime)
\label{eq:decompQ21}
\\*
Q_{22}^{c2}(12;1^\prime2^\prime)&=&Q_{22}^{c2R}(12;1^\prime2^\prime)+Q_{22}^{c2L}(12;1^\prime2^\prime)
\label{eq:decompQ22}
\\*
Q_{12}^{c1R}(1;1^\prime2^\prime) &\equiv&\nabla_R v^R(1^\prime2^\prime)\cdot\nabla_P\delta(11^\prime)
\label{eq:Q12R}
\\*
Q_{21}^{c1R}(12;1^\prime)
&\equiv& g_m\rho g^R(12)\left[{M_M(1)M_M(2)/ M_M(1^\prime)}\right]
 \nonumber
 \\*
 &&\times\nabla_Rv^R(12)\cdot\nabla_P\delta(11^\prime)
\label{eq:Q21R}
\\
Q_{22}^{c2R}(12;1^\prime2^\prime)&\equiv&\nabla_P\delta(11^\prime)\cdot\nabla_Rv^R(12)\delta(22^\prime)
\label{eq:Q22R}
\\*
Q_{12}^{c1L}(1;1^\prime2^\prime) &\equiv&\nabla_R v^L(1^\prime2^\prime)\cdot\nabla_P\delta(11^\prime)
\label{eq:Q12L}
\\
Q_{21}^{c1L}(12;1^\prime)
\nonumber
 &\equiv&\rho g^L(12)\left[M_M(1)M_M(2)/ M_M(1^\prime)\right]
 \\*
&&\times \nabla_Rv^L(12)\cdot\nabla_P\delta(11^\prime)
\label{eq:Q21L}
\\
Q_{22}^{c2L}(12;1^\prime2^\prime)&\equiv&\nabla_P\delta(11^\prime)\cdot\nabla_Rv^L(12)\delta(22^\prime)
\label{eq:Q22L}
\end{eqnarray}

\subsection{$\chi$ expressed in terms of repulsive force memory functions and scattering functions}\label{sec:repulsiveapp1}

The $Q^R$ vertices in Appendix\ \ref{sec:decompositionapp} and 
Sec.\ \ref{sec:repsrf}  represent interparticle interactions that are generated by the repulsive part of the potential of mean force.  These forces are very large when two particles or fluctuations are close together.  We can define a new set of vertices called $M^R$, $T_{12}^R$, $T_{21}^R$, and $T_{22}^R$, that can be used to replace the $Q^R$ vertices.  The new vertices represent memory functions and scattering functions for particles with short ranged repulsive forces only.

See Appendix \ref{sec:MRTRdef} for the graphical definitions of the functions that correspond to these vertices.  $M^R$ and $M_s^R$ are vertices with one left root and one right root.  The $T^R$ vertices have a number of left roots and right roots that correspond to their subscripts. The only vertices in the graphs  that define these functions are $Q_{11}^{c1}$ and $Q^R$  vertices. The only bond is $\chi^{(0)}$.  Each of the newly defined vertices has distinct time arguments for its left root(s) and right root(s). 

With these definitions, we can perform a sequence of straightforward topological reductions that eliminates the $Q^R$ vertices from the series for $\chi$  and replaces them with $M^R$ and $T^R$ vertices. The result is the following.
\vspace{2mm}

\noindent$\chi(1,t;1^{\prime},t^\prime)=$ the sum of all topologically distinct matrix diagrams with: 

\noindent$(i)$ a left root labeled $(1,t)$ and a right root labeled $(1^\prime, t^\prime)$; 

\noindent$(ii)$ free points; 

\noindent$(iii)$ $\chi^{(0)}$ bonds; 

\noindent$(iv)$ $Q_{11}^{c1}$, $Q_{11}^{c0}$,  $M^R$, $T_{12}^{R}$, $T_{21}^{R}$, $T_{22}^{R}$, $Q_{12}^{c1L}$, $Q_{21}^{c1L}$, and $Q_{22}^{c2L}$ vertices; 

\noindent such that: 

\noindent$(i)$ both roots are attached to $\chi^{(0)}$ bonds; 

\noindent$(ii)$ each free point is attached to a bond and a vertex;

\noindent$(iii)$ no pair of $T^R$ vertices are connected to each other by two particle paths containing only $\chi^{(0)}$ bonds and $Q_{11}^{c1}$ vertices.\,\,$\Box$

\subsection{The memory function and collision vertices for repulsive spheres}\label{sec:MRTRdef}

\noindent $T_{22}^R(12,t;1^\prime2^\prime,t^\prime)$

\noindent$\equiv\left(Q_{22}^{c2R}(12;1^\prime 2^\prime)+Q_{22}^{c2R}(12;2^\prime 1^\prime)\right)\delta(t-t^\prime)$ 
$+$ the sum of all topologically distinct matrix
diagrams with:

\noindent($i$) two left roots labeled $(1,t)$ and $(2,t)$ and two right roots labeled $(1^\prime, t^\prime)$ and $(2^\prime, t^\prime)$;

\noindent($ii$) free points;

\noindent($iii$) $\chi^{(0)}$ bonds;

\noindent($iv$) $Q_{11}^{c1}$ vertices and two or more $Q_{22}^{c2R}$ vertices;

\noindent such that: 

\noindent($i$) each root is on a $Q_{22}^{c2R}$ vertex;

\noindent($ii$) each free point is attached to a vertex and a bond;

\noindent($iii$) there is a particle line from root $1^\prime$ to root 1 and a particle line from root $2^\prime$ to root 2.\,\,$\Box$
\vspace{1mm}

\noindent $T_{12}^R(1,t;1^\prime 2^\prime, t^\prime)$
\noindent$\equiv Q_{12}^{c1R}(1;1^\prime 2^\prime)\delta(t-t^\prime)$ 
$+$ the sum of all topologically distinct matrix diagrams with:

\noindent$(i)$  one left root labeled $(1,t)$ and two right roots labeled $(1^\prime, t^\prime)$ and $(2^\prime, t^\prime)$;

\noindent$(ii)$  free points;

\noindent$(iii)$  $\chi^{(0)}$ bonds;

\noindent$(iv)$  $Q_{11}^{c1}$ vertices, one $Q_{12}^{c1R}$ vertex, and one or more $Q_{22}^{c2R}$ vertices;

\noindent such that:

\noindent$(i)$  the left root is on the $Q_{12}^{c1R}$ vertex and the right roots are on a $Q_{22}^{c2R}$ vertex;

\noindent$(ii)$  each free point is attached to a vertex and a bond;

\noindent$(iii)$  there is a particle line from root $1^\prime$ to root 1;

\noindent$(iv)$ there is a particle line from root $2^\prime$ to a free point on the $Q_{12}^{c1R}$.
\,\,$\Box$
\vspace{1mm}

\noindent $T_{21}^R(12,t;1^\prime, t^\prime)$
\noindent$\equiv Q_{21}^{c1R}(12;1^\prime)\delta(t-t^\prime)$ 
$+$ the sum of all topologically distinct matrix diagrams with:

\noindent$(i)$  two left roots labeled $(1,t)$ and $(2,t)$ and one right root labeled $(1^\prime, t^\prime)$;

\noindent$(ii)$  free points;

\noindent$(iii)$  $\chi^{(0)}$ bonds;

\noindent$(iv)$  $Q_{11}^{c1}$ vertices, one $Q_{21}^{c1R}$ vertex, and one or more $Q_{22}^{c2R}$ vertices;

\noindent such that: 

\noindent$(i)$  the right root is on the $Q_{21}^{c1R}$ vertex and the left roots are on a $Q_{22}^{c2R}$ vertex;

\noindent$(ii)$  each free point is attached to a vertex and a bond;

\noindent$(iii)$  there is a particle line from root $1^\prime$ to root 1;

\noindent$(iv)$ there is a particle line from a free point on the $Q_{21}^{c1R}$ vertex to root 2.
\,\,$\Box$
\vspace{1mm}

\noindent $M_s^R(1,t;1^\prime, t^\prime)\equiv$ the sum of all topologically distinct matrix diagrams with:

\noindent$(i)$  one left root labeled $(1,t)$ and one right root labeled $(1^\prime, t^\prime)$;

\noindent$(ii)$  free points;

\noindent$(iii)$  $\chi^{(0)}$ bonds;

\noindent$(iv)$  $Q_{11}^{c1}$ vertices, one $Q_{21}^{c1R}$ vertex, one $Q_{12}^{c1R}$ vertex, and $Q_{22}^{c2R}$ vertices;

\noindent such that: 

\noindent$(i)$  the left root is on the $Q_{12}^{c1R}$ vertex, and the right root is on the $Q_{21}^{c1R}$ vertex; 

\noindent$(ii)$  each free point is attached to a vertex and a bond;

\noindent$(iii)$  there is a particle line from root $1^\prime$ to root 1 and a particle line from a free point on the $Q_{21}^{c1R}$ to a free point
on the $Q_{12}^{c1R}$.
$\Box$
\vspace{1mm}

\noindent $M^R_d(1,t;1^\prime ,t^\prime)\equiv$ the sum of all topologically distinct matrix diagrams with:

\noindent$(i)$  one left root labeled $(1,t)$ and one right root labeled $(1^\prime, t^\prime)$;

\noindent$(ii)$  free points;

\noindent$(iii)$  $\chi^{(0)}$ bonds;

\noindent$(iv)$  $Q_{11}^{c1}$ vertices, one $Q_{21}^{c1R}$ vertex, one $Q_{12}^{c1R}$ vertex, and $Q_{22}^{c2R}$ vertices;

\noindent such that :

\noindent$(i)$  the left root is on the $Q_{12}^{c1R}$ vertex, and the right root is on the $Q_{21}^{c1R}$ vertex; 

\noindent$(ii)$  each free point is attached to a vertex and a bond;

\noindent$(iii)$  there is a particle line from root $1^\prime$ to a free point on the $Q_{12}^{c1R}$ and a particle line from a free point on the
$Q_{21}^{c1R}$ to root 1.
\,\,$\Box$
\vspace{1mm}

\noindent $M^R(1,t;1^\prime, t^\prime) 
\equiv M^R_s(1,t;1^\prime, t^\prime)+M^R_d(1,t;1^\prime, t^\prime)$.\quad$\Box$

\subsection{The hard sphere limit for the short ranged repulsive forces}\label{sec:hardsphere}

The only vertices in the diagrammatic expressions  for $M^R$ and $T^R$ just above in Appendix \ref{sec:MRTRdef} are $Q_{11}^{c1}$ vertices and $Q^R$ vertices.
The formulas for the $Q^R$ vertices in Eqs.\ (\ref{eq:decompQ12})-(\ref{eq:decompQ22}) for the fluid of interest are the same as those for a very low density gas of particles whose \textit {bare\,} potential is equal to $v^R$, as calculated from  Eqs.\ (\ref{eq:Q12})-(\ref{eq:Q22}), with no change except for a numerical factor of $g_m$ in the $Q_{21}^{c1R}$ vertex. Moreover, the $Q_{11}^{c1}$ vertex is of the same form for a dilute gas as for a dense liquid. This leads to the result that the $M^R$ and  $T^R$ vertices for the fluid of interest are equivalent to the $M^R$ and $T^R$ vertices of a low density gas of particles whose \textit {bare} potential is equal to $v^R$, except for an additional factor of $g_m$ in the $M^R$ and $T_{21}^R$ vertices of the fluid of interest.  (Each of these two contains one $Q_{21}^{c1R}$ vertex in each of its diagrams.)  Moreover, $M^R$ and $T^R$ describe the dynamics of just two particles colliding.  Thus the physical effects described by $M^R$ and $T^R$ are equivalent to the problem of two particles colliding with one another due to short ranged repulsions with one another.

If, for the liquid of interest, the short ranged repulsive part of the potential of mean force is very repulsive and can be approximated as a hard sphere potential, an approximation for $M^R$ and $T^R$ can be expressed in terms of the collision operator of the linearized Boltzmann equation for a dilute gas of hard sphere particles.

We want to formulate such an approximation. To do so, it is helpful to construct a formal limiting procedure by which the potential of mean force, which is a sum of a short ranged repulsive part and a longer ranged part, is smoothly converted into a sum of a hard sphere part and a longer ranged part, by variation of a parameter. See Appendix \ref{sec:conversiontohardsphere} for a discussion of how this can be done.

Then we consider what happens to the diagrammatic expressions for $M^R$ and $T^R$ as the repulsive part of the potential of mean force approaches a hard sphere potential. Before the hard sphere limit is actually reached, the diagrammatic series for the $M^R$ and $T^R$ functions are closely related to the dynamics of isolated pairs of particles interacting by short ranged, continuous repulsive forces.  As the hard sphere limit is approached, the duration of the collisions becomes smaller and smaller. In the hard sphere limit, the $Q^R$ vertices are singular for small distances between the particles, and the collisions become instantaneous, but the $M^R$ and $T^R$ vertices approach a well-defined nonsingular limit. Each vertex function can be regarded as instantaneous, with only one time argument assigned to all its points.  Moreover, the limiting vertex functions can be evaluated from the properties of hard sphere collisions. They will be denoted $M^H$ and $T^H$.    
 
Expressions for the functions associated with these vertices are given in Appendix \ref{sec:hardspherelimit}. 
The diagrammatic series for $\chi$ expressed in terms of these functions is given at the end of Sec.\ \ref{sec:repsrf}.

\subsection{A formal limiting procedure for the potential of mean force}\label{sec:conversiontohardsphere}

Consider the properties of the short ranged repulsive part of the potential of mean force that was introduced in Appendix \ref{sec:decompositionapp}.
\begin{align*}
v^R(r) &= 0 &&\text{for $r\ge r_m$}
\\*
&> 0 &&\text{for $r\le r_m$}
\\*
&\to \infty &&\text{as $r\to 0$}
\\*
dv^R(r)/dr &< 0 &&\text{for $r<r_m$}
\end{align*}
 Extend the definition of the function so that
\[v^R(r) =\infty\quad\text{for $r\le0$}\]
Let $d$ be a hard sphere diameter that is appropriate for this potential at the density of interest.  It should satisfy $d<r_m$, since $r_m$ is the range of $v^R$.
Let 
\[R(r;\epsilon{, d}) =d+(r-d)/\epsilon\]

Consider the following potential, whose arguments are two parameters, $\epsilon$ and $d$, as well as $r$.
 \[v^R(r;\epsilon,d)\equiv v^R(R(r;\epsilon,d))\]
For $\epsilon=1$
\[v^R(r;1,d)=v^R(r)\]
For $\epsilon<1$,
\begin{align*}
v^R(r;\epsilon,d) &=0&&\text{for $r\ge d+\epsilon(r_m-d)$}
\\*
&=v^R(d) &&\text{for $r=d$}
\\*
& =\infty &&\text{for $r\le d(1-\epsilon)$}
\\*
dv^R(r;\epsilon,d)/dr &< 0 &&\text{for $d(1-\epsilon) < r < d+\epsilon(r_m-d)$}
\end{align*}
This is a potential that is more repulsive than $v^R(r)$, and for small $\epsilon$, as $r$ decreases, $v^R(r;\epsilon,d)$ rises from 0 to $\infty$ continuously over a very narrow range of distances that contains $d$.  Moreover,
\[\lim_{\epsilon\to0+}v^R(r;\epsilon,d) = v^H(r;d)\]
where $v^H(r;d)$ is a potential for hard spheres with diameter $d$.
\begin{align*}
v^H(r;d) &\equiv \infty &&\text{for $r<d$}
\\*
&\equiv 0 &&\text{for $r>d$}
\end{align*}
 The effect on $v^R(r;\epsilon,d)$ of decreasing $\epsilon$ from 1 to zero is simply to convert a positive and very repulsive potential $v^R(r)$ into a hard sphere potential $v^H(r;d)$. However, for any nonzero $\epsilon$, the potential is continuous with all forces being finite at distances where the potential is finite.

\label{sec:hsapmfapp}

We assume that good approximations for the $M^R$ and $T^R$ for the fluid of interest can be obtained by choosing some appropriate value of $d$, which in general is temperature and density dependent, replacing $v^R(r)$ by $v^R(r;\epsilon,d)$, and taking the limit $\epsilon\to0$. This approximation should be a good one if in fact the repulsive part of the potential of mean force is very similar to a hard sphere potential and if the value of $d$ is chosen appropriately. In that limit, the formulas for the vertices become the same as those for a hard sphere gas at low density with one exception: The equations for $M^H$ and  $T_{21}^H$ have an additional factor of $g_m$, which is the height of the first peak of $g(r)$ for the fluid of interest.  

As a result of this assumption, the dynamics of hard sphere binary collisions can be used to approximate the effect of the short ranged repulsive $v^R$ on the dynamics of a dense liquid. See Appendix \ref{sec:hardsphere} for a discussion of the results.

This procedure requires the choice of a diameter $d$ for the hard spheres.  In the final results in the overdamped limit, the only effect of the choice of $d$ is that it determines the collision frequency $\nu$ for the hard spheres.  See Eq.\ (\ref{eq:nu}).  It is reasonable to choose $d$ such that it implies a value of $\nu$ that is the equal to the closest analogous quantity to $\nu$ for the fluid of interest.  See Sec.\ \ref{sec:discussion} for a discussion about how the value of $\nu$ can be chosen.

This assumption approximates the potential of mean force as
\[v(r)\approx v^H(r;d) +v^L(r)\]
 This is equivalent to approximating the radial distribution function as
\[g(r) = \Theta(r-d)\exp(-v^L(r)/k_BT)\]
where $\Theta$ is the Heaviside function. Since any reasonable choice of $d$ will satisfy $d<r_m$, where  $r_m$ is the distance at which the correct $g(r)$ for the liquid has its first maximum, we find that for the approximate $g$, the value of $g(r)$ at contact of the hard spheres is
\[g(d+) = \exp(-v^L(d)/k_BT) =\exp(-v_m/k_BT) = g_m\]
(See Appendix \ref{sec:decompositionapp}.)  Thus the additional factor of $g_m$ that arises in the formulas for $M^H$ and $T_{21}^H$ can be interpreted as the pair correlation function at hard sphere contact for the approximate $g(r)$.  

The approximation just described for the repulsive part of the potential of mean force, when combined with a complete neglect of the longer ranged part of the potential of mean force,  leads to results that are physically and mathematically equivalent to the generalized Enskog theory of dense hard sphere liquids, in which the atoms undergo uncorrelated binary hard sphere collisions with a collision frequency that contains a factor of the pair correlation function at contact.

\subsection{The hard sphere limit of the $M^R$ and $T^R$ vertices}\label{sec:hardspherelimit}

In the limit in which the repulsive forces become hard spheres, each $M^R$ and $T^R$ vertex has a delta function relationship between its two time arguments.  It is then convenient to regard them as instantaneous vertices that are assigned one time variable when a diagram is evaluated. The vertex functions themselves then are independent of time. 

Let $H$ (for hard sphere) rather than $R$ denote these vertices in the hard sphere limit. The derivation of the following results is very detailed and will be omitted.
  
The $T_{22}^H$ function is a scattering function that relates the position and momentum of two particles just after a hard sphere collision to their positions and momenta just before the collision.
  \begin{eqnarray*}
\lefteqn{T_{22}^H(12;1^{\prime}2^{\prime})}
\\*
&=& \Theta(-{\mathbf P}_{12}^{\prime}\cdot{\mathbf R}_{12}^{\prime})|(1/m){\mathbf P}_{12}^{\prime}\cdot\hat{\mathbf R}_{12}^{\prime}|\delta(|{\mathbf R}_{12}^{\prime}|-d)
\\*
&&\times[\delta({\mathbf P}_1-{\mathbf p}_{1}({1}^{\prime}2^{\prime}))\delta({\mathbf P}_2
-{\mathbf
p}_{2}(1^{\prime}2^{\prime}))
\\*
&&\quad-\delta({\mathbf P}_{1}-{\mathbf P}_1^{\prime})\delta({\mathbf P}_{2}-{\mathbf
P}_2^{\prime})]
\\*
&&\times\, \delta({\mathbf R}_1-{\mathbf R}_1^{\prime})
\delta({\mathbf R}_2-{\mathbf R}_2^{\prime})
\\*
&=& \Theta({\mathbf P}_{12}\cdot{\mathbf R}_{12})|(1/m){\mathbf P}_{12}\cdot\hat{\mathbf R}_{12}|\delta(|{\mathbf R}_{12}|-d)
\\*
&&\times[\delta({\mathbf P}_1^\prime-{\mathbf p}_{1}(1{2}))
\delta({\mathbf P}_2^\prime-{\mathbf p}_{2}(12))
\\*
&&\quad-\delta({\mathbf P}_{1}-{\mathbf P}_1^{\prime})\delta({\mathbf P}_{2}-{\mathbf
P}_2^{\prime})]
\\*
&&\times\, \delta({\mathbf R}_1-{\mathbf R}_1^{\prime})
\delta({\mathbf R}_2-{\mathbf R}_2^{\prime})
\end{eqnarray*}
Here
\begin{eqnarray*}
          {\mathbf p}_{1}(12) &=& {\mathbf P}_1-({\mathbf P}_{12}\cdot\hat{\mathbf R}_{12})\hat{\mathbf R}_{12}
\\*
{\mathbf p}_{2}(12) &=& {\mathbf P}_2+({\mathbf P}_{12}\cdot\hat{\mathbf R}_{12})\hat{\mathbf R}_{12}
\end{eqnarray*}
and 
\begin{align*}
{\mathbf P}_{12} &={\mathbf P}_1-{\mathbf P}_2
\\*
{\mathbf R}_{12} &= {\mathbf R}_1-{\mathbf R}_2
\\*
\hat{\mathbf R}_{12} &= {\mathbf R}_{12}/|{\mathbf R}_{12}|
\end{align*}
 The other functions are directly related to $T_{22}^H$.
\begin{eqnarray*}
T_{21}^H(12;1^\prime) &=&\int d2^\prime\,T_{22}^H(12;1^\prime2^\prime)\rho g_m M_M(2^\prime)
\\
T_{12}^H(1;1^\prime 2^\prime) &=&\int d2\,T_{22}^H(12;1^\prime2^\prime)
\\
M_s^H(1;1^\prime) &=&\int d2\,T_{21}^H(12;1^\prime )
\\*
& =&\int d2d2^\prime\,T_{22}^H(12;1^\prime2^\prime)\rho g_m M_M(2^\prime)
\\*
M_d^H(1;1^\prime) &=&\int d2\,T_{21}^H(21;1^\prime)
\\*
&=&\int d2d2^\prime\,T_{22}^H(21;1^\prime2^\prime)\rho g_m M_M(2^\prime)
\\
M^H(1;1^\prime) &=&M_s^H(1;1^\prime)+M_d^H(1;1^\prime)
\end{eqnarray*}

$T_{22}^H$ is a vertex with two left points and two right points.  There is an internal line between the first right point and the first left point and another between the second right point and second left point.  The function for this vertex, as well as the symbol for the vertex used in graphs, is symmetric under the simultaneous interchanges of $1\leftrightarrow2$ and $1^\prime\leftrightarrow2^\prime$, but in general this does not lead to any symmetry number considerations for the graphs used in this theory.

$T_{12}^H$ is a vertex with one left point and two right points.  There is an internal line between the first right point and the left point.

$T_{21}^H$ is a vertex with two left points and one right point.  There is an internal line between the right point and the first left point.

$M_s^H$ has an internal line between its points.  $M_d^H$ has no internal line. 

\section{Details of the Hermite polynomial representation}\label{sec:Hermite}

\subsection{Construction of the Hermite representation}

The functions $H_n(x)$ in Eq.\ (\ref{eq:specialHermitepolynomial}) are Hermite polynomials of the standard variety defined by the following equation.
\begin{equation*}
H_{n}(x)=(-1)^{n}e^{x^{2}}
\frac{d^{n}}{dx^{n}}e^{-x^{2}}
\end{equation*}
The special Hermite polynomials obey the following orthogonality relation.
\begin{equation*}
\int d\mathbf{P}\, h_{\lambda}(\mathbf{P})h_{\lambda^{'}}(\mathbf{P})M_M(\mathbf{P})=
\delta(\lambda,\lambda^{'})
\end{equation*}
where the function on the right is a Kronecker delta function.
 The above relation can be used to solve Eq.\ \eqref{eq:chiHermite} for $\chi(\mathbf{R},\lambda,t;\mathbf{R}^\prime,\lambda^\prime,t^\prime)$.
\begin{eqnarray*}
\lefteqn{\chi(\mathbf{R},\lambda,t;\mathbf{R}^\prime,\lambda^{\prime},t^\prime)}
\\*
&=&\int d\mathbf{P} d\mathbf{P}^\prime\, h_{\lambda}(\mathbf{P})\chi(\mathbf{R},\mathbf{P},t;\mathbf{R}^\prime,\mathbf{P}^\prime,t^\prime)
h_{\lambda^\prime}(\mathbf{P}^{\prime})
\end{eqnarray*}
We refer to this quantity as the $(\lambda,\lambda^{\prime})$ matrix element of $\chi$.  Using analogous relationships, it is possible to define and calculate the matrix elements of the functions associated with individual vertices and bonds. The results are given in Appendix \ref{sec:matrixelements}. Just as the function $\chi$ itself is described by a diagrammatic series containing certain bonds and vertices, the Hermite matrix elements of $\chi$ are described by a topologically identical series whose diagrams are evaluated using the Hermite matrix elements of the original bond functions and vertex functions. 

\subsection{Evaluation of diagrams in the Hermite representation}\label{sec:Hermiteevaluationapp}

To evaluate a diagram in this series, use the following prescription: 

\noindent$(i)$ assign to each instantaneous vertex in the diagram a dummy time argument and to each retarded vertex a dummy time argument for its right points and a dummy time argument  for its left points;

\noindent$(ii)$ assign to each free point a dummy position argument, a dummy Hermite index, and the dummy time argument of the vertex to which it is attached;  

\noindent $(iii)$ construct a summand consisting of the product of the Hermite matrix elements of the functions for each vertex and bond in the diagram, with the Hermite arguments being the dummy variables assigned to the free points and the variables assigned to each root;  

\noindent$(iv)$ sum the summand over all the dummy Hermite indices, and integrate the sum over all values of the dummy position and time arguments.  

The result is the value of the diagram, which is a function of the positions, Hermite indices, and times associated with the roots.

As an example, consider the fourth diagram in Fig.\ \ref{fig:chi} (i.e.\ the diagram on the right in the second line).  In the Hermite representation, the left root is labeled $({\mathbf R},{\mathbf \lambda},t)$ and the right root is labeled $({\mathbf R}^\prime,{\mathbf \lambda}^\prime,t^\prime)$. 

 \noindent$(i)$ Assign the dummy time arguments $t_1$ to the vertex on the left and $t_2$ to the vertex on the right. 
 
\noindent$(ii)$ Assign labels $i=1,\ldots,6$ to the six free points.  For example, assign 1 to the free point on the right of the right vertex, 2 to the free point on the upper left of that vertex, 3 to the lower free point on the left of vertex, 4 to the upper right point on the other vertex, 5 to the lower right point on that vertex, and 6 to the left point on that vertex.  Associated with each label $i$ is a dummy position argument ${\mathbf R}_i$ and a dummy Hermite index $\lambda_i$.

\noindent$(iii)$ For each vertex and bond, assign arguments to its function that corresponds to the way that variables have been assigned to its points. 

  As an example of a vertex, consider the vertex on the right.  It is a $Q_{21}^{c1}$ vertex.  Its right point is 1, the point on the left that is connected internally to 1 is 3, and the other point on the left is 2.  Therefore the function associated with this vertex  is
\[Q_{21}^{c1}({\mathbf R}_3,\lambda_3,{\mathbf R}_2,\lambda_2;{\mathbf R}_1,\lambda_1)\]
The convention is that the first left argument and the right argument (i.e.\ 3 and 1, respectively) correspond to the two points connected by an internal line.  This vertex function is not time dependent, so there are no time arguments in this.

As an example of a bond, consider the upper bond in the middle of the diagram.  It is a $\chi^{(0)}$ bond.  Its left point is 4 and its right point is 2.  The time argument of the left point is $t_2$ and that for the right point is $t_1$.  Therefore the function associated with this bond is
\[\chi^{(0)}({\mathbf R}_4,\lambda_4,t_2;{\mathbf R}_2,\lambda_2,t_1)\]

The summand is the product of all such functions for all four bonds and two vertices in the diagram.  This is then summed over all dummy Hermite indices and integrated over all values of the dummy position and time variables.
\begin{align*}
\lefteqn{\text{the value of the diagram} =
}
\\*
&\times\int_{-\infty}^\infty dt_1\int_{-\infty}^\infty dt_2
\sum_{\lambda_1\ldots\lambda_6}\int d\mathbf R_1\ldots d\mathbf R_6d\mathbf P_1\ldots d\mathbf P_6
\\*
&\times
\chi^{(0)}({\mathbf R},\lambda,t;{\mathbf R}_6,\lambda_6,t_2)
Q_{12}^{c1}({\mathbf R}_6,\lambda_6;{\mathbf R}_4,\lambda_4,{\mathbf R}_5,\lambda_5)
\\*
&\times\chi^{(0)}({\mathbf R}_4,\lambda_4,t_2;{\mathbf R}_2,\lambda_2,t_1)
\chi^{(0)}({\mathbf R}_5,\lambda_5,t_2;{\mathbf R}_3,\lambda_3,t_1)
\\*
&\times Q_{21}^{c1}({\mathbf R}_3,\lambda_3,{\mathbf R}_2,\lambda_2;{\mathbf R}_1,\lambda_1)
\chi^{(0)}({\mathbf R}_1,\lambda_1,t_1;{\mathbf R}^\prime,\lambda^\prime,t^\prime)
\end{align*}
\subsection{Matrix elements of the bonds and vertices} \label{sec:matrixelements}
\subsubsection{Bonds and vertices of the original theory}
 \begin{eqnarray} 
\lefteqn{
\chi^{(0)}(\mathbf{R}_1,\lambda_1,t;\mathbf{R}^{'}_{1},\lambda^{'}_{1},t')
}
\nonumber
\\*
&=&\delta(\mathbf{R}_1-\mathbf{R}_{1}^{'})\Theta(t-t')\delta(\lambda_1,\lambda_{1}^{'})
\label{eq:chizerodef}
\\*
\lefteqn{Q_{11}^{c1}(\mathbf{R}_1,\lambda_1;\mathbf{R}^{'}_{1},\lambda^{'}_{1})}
\nonumber
\\*
&=&-v_T\sum_{i=x,y,z}\partial_{R_{1i}}\delta(\mathbf{R}_1-\mathbf{R}_{1}^{'}) \nonumber 
\\*
&\times&\left((\lambda_{1i}+1)^{1/2}\delta(\lambda_{1i}+1,\lambda_{1i}^{'})+
(\lambda_{1i})^{1/2}\delta(\lambda_{1i}-1,\lambda_{1i}^{'})\right)
\nonumber
\\*
&&\times\prod_{j(\neq i)=x,y,z}
\delta(\lambda_{1j},\lambda_{1j}^{'})
\label{eq:Q11c1def}
\end{eqnarray}
Here, $v_T$ is the thermal velocity $(k_BT/m)^{1/2}$.
\begin{eqnarray}
\label{eq:Q11c0def}
\nonumber
\lefteqn{Q_{11}^{c0}(\mathbf{R}_1,\lambda_1;\mathbf{R}^{'}_{1},\lambda^{'}_{1})}
\\*
&=&v_T\rho\delta(\lambda_{1}^{'},\hat{0})\sum_{i=x,y,z}\partial_{R_{1i}}c(\mathbf{R}_1-\mathbf{R}_{1}^{'}) \nonumber 
\\*
&&\times (\lambda_{1i})^{1/2}\delta(\lambda_{1i}-1,\hat{0})\prod_{j(\neq i)=x,y,z}
\delta(\lambda_{1j},\hat{0})\quad\quad
\end{eqnarray}
\begin{eqnarray}
\lefteqn{Q_{12}^{c1}(\mathbf{R}_1,\lambda_1;\mathbf{R}_{1}^{'},\lambda_{1}^{'},\mathbf{R}_{2}^{'},\lambda_{2}^{'})}
\nonumber
\\*
&=&-(mv_T)^{-1}\delta(\lambda_{2}^{'},\hat{0})\delta(\mathbf{R}_1-\mathbf{R}_{1}^{'}) \nonumber \\
&\times&\sum_{i=x,y,z}\partial_{R_{1i}}v(\mathbf{R}_1-\mathbf{R}_{2}^{'})
(\lambda_{1i})^{1/2}
\nonumber
\\*
&&\times\delta(\lambda_{1i}-1,\lambda_{1i}^{'})
\prod_{j(\neq i)=x,y,z}\delta(\lambda_{1j},\lambda_{1j}^{'})
\label{eq:Q12def}
\\*
\lefteqn{Q_{21}^{c1}(\mathbf{R}_1,\lambda_1,\mathbf{R}_2,\lambda_2;\mathbf{R}_{1}^{'},\lambda_{1}^{'})}
\nonumber
\\*
&=&-v_T\rho \delta(\mathbf{R}_1-\mathbf{R}_{1}^{'})\delta(\lambda_2,\hat{0}) \nonumber \\
&&\times\sum_{i=x,y,z}\partial_{R_{1i}^{'}}\exp\left(-v(\mathbf{R}_{1}^{'}-\mathbf{R}_2)/k_BT\right)
(\lambda_{1i}^{'})^{1/2}
\nonumber
\\*
&&\quad\times\delta(\lambda_{1i}^{'}-1,\lambda_{1i})
\prod_{j(\neq i)=x,y,z}\delta(\lambda_{1j}^{'},\lambda_{1j})
\label{eq:Q21def}
\\*\lefteqn{Q_{22}^{c2}(\mathbf{R}_1,\lambda_1,\mathbf{R}_2,\lambda_2;
\mathbf{R}_{1}^{'},\lambda_{1}^{'},\mathbf{R}_{2}^{'},\lambda_{2}^{'})}
\nonumber
\\*
&=&-(mv_T)^{-1}\delta(\mathbf{R}_1-\mathbf{R}_{1}^{'})\delta(\mathbf{R}_2-\mathbf{R}_{2}^{'})
\delta(\lambda_2,\lambda_{2}^{'}) \nonumber \\
&&\times\sum_{i=x,y,z}\partial_{R_{1i}}v(\mathbf{R}_1-\mathbf{R}_{2})
(\lambda_{1i})^{1/2}\delta(\lambda_{1i}-1,\lambda_{1i}^{'})
\nonumber
\\*
&&\times\prod_{j(\neq i)=x,y,z}\delta(\lambda_{1j},\lambda_{1j}^{'})
\label{eq:Q22def}
\end{eqnarray}

\subsubsection{Hard sphere vertices}
\label{sec:hardsphereHermite}
\begin{eqnarray}
\lefteqn{T_{22}^H({\mathbf R}_1,\lambda_1,{\mathbf R}_2,\lambda_2;{\mathbf R}_1^\prime,\lambda_1^\prime,{\mathbf R}_2^\prime,\lambda_2^\prime)}
\nonumber
\\*
&=&\int d{\mathbf P}_1d{\mathbf P}_2\,
h_{\lambda_1}({\mathbf P}_1)
h_{\lambda_2}({\mathbf P}_2)
\nonumber
\\*
&&\times\left[h_{\lambda_1^\prime}({\mathbf p}_{1}(1{2}))
h_{\lambda_2^\prime}({\mathbf p}_{2}(12))
-h_{\lambda_1^\prime}({\mathbf P}_{1})
h_{\lambda_2^\prime}({\mathbf P}_2)\right]
\nonumber
\\
&&\times M_M({\mathbf P}_1)
M_M({\mathbf P}_2)
\Theta({\mathbf P}_{12}\cdot{\mathbf R}_{12})|(1/m){\mathbf P}_{12}\cdot\hat{\mathbf R}_{12}|
\nonumber
\\*
&&\times
 \delta(|{\mathbf R}_{12}|-d)\delta({\mathbf R}_1-{\mathbf R}_1^{\prime})
\delta({\mathbf R}_2-{\mathbf R}_2^{\prime})
\nonumber
\\
&=&\int d{\mathbf P}_1^\prime d{\mathbf P}_2^\prime\,
\Theta(-{\mathbf P}_{12}^{\prime}\cdot{\mathbf R}_{12}^{\prime})|(1/m){\mathbf P}_{12}^{\prime}\cdot\hat{\mathbf R}_{12}^{\prime}|
\nonumber
\\*
&&\times\left[
h_{\lambda_1}({\mathbf p}_{1}({1}^{\prime}2^{\prime}))
h_{\lambda_2}({\mathbf
p}_{2}(1^{\prime}2^{\prime}))
-h_{\lambda_1}({\mathbf P}_1^{\prime})
h_{\lambda_2}({\mathbf
P}_2^{\prime})
\right]
\nonumber
\\*
&&\times
h_{\lambda_1^\prime}({\mathbf P}_1^\prime)
h_{\lambda_2^\prime}({\mathbf P}_2^\prime)
M_M({\mathbf P}_1^\prime)
M_M({\mathbf P}_2^{\prime})
\nonumber
\\*
&&\times\delta(|{\mathbf R}_{12}^{\prime}|-d) \delta({\mathbf R}_1-{\mathbf R}_1^{\prime})
\delta({\mathbf R}_2-{\mathbf R}_2^{\prime})
\nonumber
\\*
\lefteqn{T_{12}^H({\mathbf R}_1,\lambda_1;{\mathbf R}_1^\prime,\lambda_1^\prime,{\mathbf R}_ 2^\prime,\lambda_2^\prime)}
\nonumber
\\*
&=&\int d{\mathbf P}_1d{\mathbf P}_1^\prime d{\mathbf P}_2^\prime\,M({\mathbf P}_1)M({\mathbf P}_1^\prime)M({\mathbf P}_2^\prime)
\nonumber
\\*
&&\times h_{\lambda_1}({\mathbf P}_1)
h_{\lambda_1^\prime}({\mathbf P}_1^\prime)
h_{\lambda_2^\prime}({\mathbf P}_2^\prime)
\nonumber
\\
&&\times\Theta(-{\mathbf P}_{12}^{\prime}\cdot{\mathbf R}_{12}^{\prime})|(1/m){\mathbf P}_{12}^{\prime}\cdot\hat{\mathbf R}_{12}^{\prime}|\delta(|{\mathbf R}_{12}^{\prime}|-d)
\nonumber
\\*
&&\times\left[\delta({\mathbf P}_1-{\mathbf p}_{1}({1}^{\prime}2^{\prime}))-\delta({\mathbf P}_{1}-{\mathbf P}_1^{\prime})\right]
 \delta({\mathbf R}_1-{\mathbf R}_1^{\prime})
\nonumber
\end{eqnarray}
\begin{eqnarray}
\lefteqn{T_{21}^H({\mathbf R}_1,\lambda_1,{\mathbf R}_2,\lambda_2;{\mathbf R}_1^\prime,\lambda_1^\prime)}
\nonumber
\\*
&=&\rho g_m
T_{22}^H({\mathbf R}_1,\lambda_1,{\mathbf R}_2,\lambda_2;
{\mathbf R}_1^\prime,
\lambda_1^\prime,
{\mathbf R}_2^\prime,\hat0)
\nonumber
\\*\lefteqn{M_s^H({\mathbf R}_1,\lambda_1;{\mathbf R}_1^\prime,\lambda_1^\prime)}
\nonumber
\\*
&=&\rho g_m\int d{\mathbf R}_2d{\mathbf R}_2^\prime\,
T_{22}^H({\mathbf R}_1,\lambda_1,{\mathbf R}_2,\hat0;{\mathbf R}_1^\prime,\lambda_1^\prime,{\mathbf R}_2^\prime, \hat0)
\nonumber
\end{eqnarray}
\begin{eqnarray}
\lefteqn{M_d^H({\mathbf R}_2,\lambda_2;{\mathbf R}_1^\prime,\lambda_1^\prime)}
\nonumber
\\*
&=&\rho g_m\int d{\mathbf R}_1 d{\mathbf R}_2^\prime\,
T_{22}^H({\mathbf R}_1,\hat0,{\mathbf R}_2,\lambda_2;{\mathbf R}_1^\prime,\lambda_1^\prime,{\mathbf R}_2^\prime,\hat0)
\nonumber
\\*
\lefteqn{\hat M_s^H({\mathbf q})_{\hat z\hat z}}
\\*
&=&-\frac{\rho g_m}{6mk_BT}\int d{\mathbf R}_2
d{\mathbf P}_1 d{\mathbf P}_2\,
\Theta(-{\mathbf P}_{12}\cdot{\mathbf R}_{12})
\nonumber
\\*
\lefteqn{\quad\times|(1/m){\mathbf P}_{12}\cdot\hat{\mathbf R}_{12}|
\delta(|{\mathbf R}_{12}|-d)
({\mathbf P}_{12}\cdot\hat{\mathbf R}_{12})^2}
\nonumber
\\*
\lefteqn{\quad\times M_M({\mathbf P}_1)
M_M({\mathbf P}_2)}
\nonumber
\\*
&=&
-\nu
\label{eq:nu}
\end{eqnarray} 
The quantity $\nu$ has dimensions of (time)$^{-1}$ and is clearly positive.  Note that $\hat{M}_s^H({\mathbf q})$ does not depend on ${\mathbf q}$ because the real space version of the function is proportional to a Dirac delta function of the difference in its position arguments.

\textbf{Nonzero matrix elements.}
Using the formulas above, it is possible to prove that a matrix element of $M^H$, $M_s^H$, $M_d^H$, $T_{22}^H$, $T_{12}^H$, or $T_{21}^H$  is nonzero only if at least one left index is not $\hat0$ and at least one right index is not $\hat0$.

\section{Graphical analysis and results}\label{sec:graphicalconcepts}

\subsection{Quasi-simultaneity of objects}\label{sec:qsdef}

See Sec.\ \ref{sec:background2} for the definition of a path.

\textbf{Definition.}  A quasi-forward path between two objects in a diagram is a path such that the tail end of each $\chi_{\hat0}^{(0)}$ bond in the path is attached to the object that precedes it in the path, and the head end of every $\chi_{\hat0}^{(0)}$ bond in the path is attached  to the object that follows it in the path, but there is no similar restriction on the $\chi_P^E$ bonds in the path.\,\,$\Box$

When a graph is evaluated, a dummy integration time variable is assigned to each vertex. The integrand of the integral contains factors for each bond that are retarded functions of its time arguments. As a result, the integrand is nonzero only when the dummy variables are consistent with the retarded nature of the bonds.  Along a path of the type just described, each vertex and root has a time variable. Because of the last restriction in the definition, when the integrand is nonzero, the time variable increases as we pass from a vertex to another vertex along a $\chi_{\hat0}^{(0)}$ bond. However, when the path passes along a $\chi_{P}^{E}$ bond there is no such restriction.  When it passes along a $\chi_{P}^{E}$ bond, the time variable can move forward or backward, but only for a very small change in time, of $O(\nu^{-1}$), if the integrand is nonzero.  As a result, if there is a quasi-forward path between two vertices and/or roots,  the integrand of the diagram is nonzero only when each time variable associated with a vertex or root on the path is greater than or  equal to the time on the vertex or root at the start of the path or precedes that time by no more than an amount of $O(\nu^{-1})$.

\textbf{Definition.}  Two objects are quasi-simultaneous (qs) if there is a quasi-forward path from the first to the second as well as a quasi-forward path from the second to the first.\,\,$\Box$

If two vertices or roots in a diagram satisfy this definition, then it is clear that the integrand of the diagram is nonzero only when the difference between the two time arguments has a magnitude of $O(\nu^{-1})$, hence the term `quasi-simultaneous' is appropriate.

This definition is equivalent to the one used in Sec.\ \ref{sec:qs}.

\subsection{Proof of a lemma used in Sec.\ \ref{sec:qs}}
\label{sec:atleasttwovertices}

\textbf{Lemma.} Every maximal qs subgraph in a diagram has

\noindent 1.\quad at least one object that has no $\chi_P^E$ bond on the left,

\noindent 2.\quad at least one object that has no $\chi_P^E$ bond on the right,

\noindent 3.\quad at least two objects that are not  $T_{21}^H$ vertices.

\textbf{Proof.}  1.  Given a maximal qs subgraph,  pick one object in the subgraph.  If it has no $\chi_P^E$ on the left, then statement 1 holds for the subgraph.  If it has a $\chi_P^E$ on the left, move along that bond to the object at the other end of that bond.  (This is motion along a forward path.) That object is in the maximal qs subgraph.  If this second object has no $\chi_P^E$ bond on the left, then statement 1 holds for the subgraph.  If not, continue this process of moving along a forward path consisting of $\chi_P^E$ bonds.  If no vertex with no $\chi_P^E$ bonds on the left is encountered, eventually the forward path must lead to the left root, which is an object with no $\chi_P^E$ object on the left.  (See Appendix \ref{sec:validgraphs}.) Therefore statement 1 holds for the subgraph.

2.  A similar argument holds for statement 2.

3.  Neither of the two objects in statements 1 and 2 can be a  $T_{21}^H$ vertex because each $T_{21}^H$ vertex must have a $\chi_P^E$ on the left and a $\chi_P^E$ on the right.  Therefore statement 3 holds.

\subsection{Proof of a theorem used in Sec.\ \ref{sec:chardiagret}}
\label{sec:majorproof}

In this appendix, except for the statement of the last theorem, all diagrams referred to are those in the series for  $\chi_P$ in Sec.\ \ref{sec:chiEQLTH}, but this will not be explicitly stated below.  The last theorem then makes a statement about the series for $\chi_P$ in the overdamped limit.

\textbf{Theorem.} Every vertex in a diagram  satisfies the overdamped vertex restrictions \textit {if and only if}\ every maximal $\chi_P^E$-connected subgraph in the diagram is an overdamped subgraph.\,\,$\Box$

The `\textit {if}\,' part of the theorem follows very directly from the lemma in Sec.\ \ref{sec:propovsub}. The `\textit {only if}\,' part is proven by detailed consideration of how the overdamped vertex restrictions limit the types of maximal $\chi_P^E$-connected subgraphs that can be constructed.

\textbf{Lemma.}\ In an overdamped subgraph, there is only one vertex with a right point that has no bond attached, namely the object usually drawn at the right of the subgraph. If this object is a root, then no vertex in the subgraph has the left point of a $\chi_{\hat0}^{(0)}$ bond attached in the graph.  If this object is a vertex, then the right point or points of this vertex are the only points in the subgraph to which the left point of a $\chi_{\hat0}^{(0)}$ bond can be attached in the graph.\,\,$\Box$
\quad The proof follows from detailed consideration of the definition of an overdamped subgraph.

\textbf{Lemma.}  If every maximal $\chi_P^E$-connected subgraph in a diagram is an overdamped subgraph, then every maximal $\chi_P^E$-connected subgraph is a maximal qs subgraph.

\textbf{Proof.} The proof is by contradiction.  

Assume that there is a maximal $\chi_P^E$-connected subgraph that is not a maximal qs subgraph. Call it $A$. Then it must be qs with at least one other $\chi_P^E$-connected subgraph, which will be called $B$.

This implies that there is a quasi-forward path from $A$ to $B$ in the diagram and a quasi-forward path from $B$ to $A$.  (See Sec.\ \ref{sec:qsdef}.)  The quasi-forward path from $A$ to $B$ must leave $A$ from some point on the left of a vertex in $A$.  This departure takes place along a $\chi_{\hat0}^{(0)}$ bond.  The path must eventually arrive at the vertex along a $\chi_{\hat0}^{(0)}$ bond on the right of the right vertex of $B$.  (See the lemma just above.  The vertex on the right of an overdamped subgraph is the only vertex on that subgraph to which the head of a $\chi_{\hat0}^{(0)}$ can be attached.)  A similar statement can be made about the quasi-forward path from $B$ to $A$.  Also, if the path visits another maximal $\chi_P^E$-connected subgraph other than $A$ or $B$, it must enter that subgraph at the vertex on the right and must leave that subgraph at the left of some vertex in the subgraph.

For any of these quasi-forward paths that enters and then leaves a $\chi_P^E$ subgraph, it is possible to adjust the path so that it always moves forward in time along $\chi_P^E$ bonds.  As a result, there exists a forward path from $A$ to $B$.  Similarly, there is a forward path from $B$ to $A$.

Finally by connecting these paths with forward paths along $\chi_P^E$ bonds in the interior of $A$ and $B$, we can show that there exists a forward path from the right object in $A$ to itself.  But this contradicts one of the fundamental properties of the diagrams that appear in this theory.  (See Sec.\ \ref{sec:background2}.)  Q.E.D.\,\,$\Box$

\textbf{Theorem.}  If every maximal $\chi_P^E$-connected subgraph in a diagram is an overdamped subgraph, every maximal qs subgraph in the diagram is a maximal $\chi_P^E$-connected subgraph.

\textbf{Proof.} The previous theorem implies that, given the assumption stated in the theorem, the number of maximal qs subgraphs in a diagram is no less than the number of maximal $\chi_P^E$-connected subgraphs in the diagram.  But every maximal qs subgraph must contain at least one maximal $\chi_P^E$-connected subgraph.  Hence the number of maximal qs subgraphs is equal to the number of maximal $\chi_P^E$-connected subgraphs.  It follows from the previous theorem that every maximal $\chi_P^E$-connected subgraph is a maximal qs subgraph.  Hence every maximal qs subgraph in the diagram is a maximal $\chi_P^E$-connected subgraph.  Q.E.D.\,\,$\Box$.

\textbf{Corollary.}   If every maximal $\chi_P^E$-connected subgraph in a diagram is an overdamped subgraph, every maximal qs subgraph in the diagram is an overdamped subgraph.\,\,$\Box$

\textbf{Theorem.}  If every vertex in a diagram satisfies the overdamped vertex restrictions, then every maximal qs subgraph in the diagram is an overdamped subgraph.  The diagram is included in the series for $\chi_P$ for large $\nu$.

\textbf{Proof.} The first part of the theorem follows directly from the first theorem of this section and corollary just above.  The second part of the theorem follows from the first part and the theorem in Sec.\ \ref{sec:thmaxqs}.\,\,$\Box$

This theorem is the main result of this subsubsection.  It is used in Sec.\ \ref{sec:chardiagret}.

\section{The calculation of  correlation functions}\label{sec:calculateobservables}
In this appendix, we summarize the formulas needed to calculate numerical results for the correlation functions of interest using approximate numerical results for $\hat m_{irr}(\mathbf q,t)_{\lambda\lambda^\prime}$, $\hat{\chi}_P^{ED}({\bf q})_{\lambda\lambda^\prime}$, and their self functions obtained in  separate calculations. 

The series for the overdamped limit of $\chi_P$ in Sec.\ \ref{sec:chiP0dDyson} has a simple geometric series representation in the Fourier-Laplace domain.
\[\hat{\tilde\chi}_P({\mathbf q},z) =\hat{\chi}_P^{ED}({\mathbf q})\left({\mathbf I}+\sum_{n=1}^\infty\left[\hat{\tilde m}_{irr}({\mathbf q},z)\hat{\chi}_P^{ED}({\mathbf q})\right]^n\right)\]
This is a Hermite matrix equation, ${\mathbf I}$ is the identity matrix, and all multiplications on the right are matrix multiplications.

For convenience in performing numerical calculations, we define an irreducible memory function as the function between square brackets in the power series above.
\begin{eqnarray*}
\hat{\tilde M}_{irr}({\mathbf q},z) &\equiv&\hat{\tilde m}_{irr}({\mathbf q},z)\hat{\chi}^{ED}_{P}({\mathbf q})
\end{eqnarray*}
The numerical values of $\hat M_{irr}({\mathbf q},t)$ can easily be calculated from the time domain version of this equation.  (In practice, the approximations  made for the matrix elements of $\hat m_{irr}({\mathbf q},t)$ will be such that $\hat{\tilde M}_{irr}(\mathbf q,z)$ calculated using this formula has a finite number of nonzero matrix elements.  As a result, all subsequent matrix calculations will in effect involve finite dimensional square matrices.)

We also define a reducible memory function as the sum in the expression above.
\begin{eqnarray*}
\hat{\tilde M}_{red}({\mathbf q},z) &\equiv&\sum_{n=1}^\infty\left[\hat{\tilde M}_{irr}({\mathbf q},z)\right]^n
\end{eqnarray*}
This implies
\begin{eqnarray*}
\hat{\tilde M}_{red}({\mathbf q},z) =\hat{\tilde M}_{irr}({\mathbf q},z)+\hat{\tilde M}_{irr}({\mathbf q},z)\hat{\tilde M}_{red}({\mathbf q},z)
\end{eqnarray*}
In the time domain, this is an integral equation that can be used to calculate $M_{red}$ from $M_{irr}$.
Then we have
\begin{eqnarray}
\hat{\tilde\chi}_{P}({\mathbf q},z) &=&\hat{\chi}^{ED}_{P}({\mathbf q})+\hat{\chi}^{ED}_{P}({\mathbf q})\hat{\tilde M}_{red}({\mathbf q},z)
\label{eq:chiPnum}
\end{eqnarray}
which can be used to calculate the numerical results for $\hat{\chi}_P({\mathbf q},t)$.

The memory function equation for the matrix element of the propagator that is closely related to the coherent intermediate scattering function is given in Eq.\ (\ref{eq:dchi00dt}), with the memory function expressed in terms of $\chi_P$ in Eq.\ (\ref{eq:MitochiP}).
Combining these equations with the latest result for $\chi_P$, we get an expression for the memory function in the overdamped limit. 
\begin{eqnarray*}
{\hat{\tilde M}({\mathbf q},z)}
& \to&\hat{\mathcal Q}_{11}({\mathbf q})+\hat{\tilde M}{}^{O}({\mathbf q},z)
\end{eqnarray*}
where
\begin{equation}
\hat{\mathcal Q}_{11}({\mathbf q}) \equiv
\left[\hat Q_{11}^{c1}({\mathbf q})\hat{\chi}{}^{ED}_{P}({\mathbf q})(\hat Q_{11}^{c1}({\mathbf q})+\hat Q_{11}^{c0}({\mathbf q}))\right]_{\hat0\hat0}
\label{eq:calQ}
\end{equation}
\begin{eqnarray}
\lefteqn{\hat{\tilde M}{}^{O}({\mathbf q},z)}
\nonumber
\\*
 &\equiv&\left[\hat Q_{11}^{c1}({\mathbf q})\hat{\chi}{}^{ED}_{P}({\mathbf q})\hat{\tilde M}_{red}({\mathbf q},z)(\hat Q_{11}^{c1}({\mathbf q})+\hat Q_{11}^{c0}({\mathbf q}))\right]_{\hat0\hat0}\quad\quad
\label{eq:MO}
\end{eqnarray}
These two functions can be calculated numerically from the previous numerical results using these formulas.
The first part of the limiting result, $\hat{\mathcal Q}_{11}({\mathbf q})$, is independent of $z$, so in the time domain it is proportional to a Dirac delta function of time.  Such a contribution is usually not regarded as part of the memory function.  The second  part,  $\hat{\tilde M}{}^{O}({\mathbf q},z)$, in the time domain is a continuous function of time, and we shall call it the overdamped memory function.  

In the overdamped limit, Eq.\ (\ref{eq:dchi00dt}) becomes
\begin{eqnarray}
\lefteqn{\frac{\partial \hat\chi(q\hat{\mathbf k},t)_{\hat0\hat0}}{\partial t}=\delta(t)+\hat{\mathcal Q}_{11}(q\hat{\mathbf k})\hat\chi(q\hat{\mathbf k},t)_{\hat0\hat0}}
\nonumber
\\&&\quad\quad\quad+
\int_0^t dt^{\prime}\,\hat M^O(q\hat{\mathbf k},t-t^\prime)\hat\chi({\mathbf q},t^{\prime})_{\hat0\hat0}
\label{eq:dchi00dt2}
\end{eqnarray}
which is the overdamped limit memory function equation for the coherent intermediate scattering function.  See Eq.\ (\ref{eq:phirhodef}).  
Numerical solution of this equation in the time domain gives results for the coherent intermediate scattering function.

To calculate the current correlation functions (see Eqs.\ (\ref{eq:phijldef}), (\ref{eq:phijtdef}), and (\ref{eq:phijlsdef})) we use the general expressions in Eq.\ (\ref{eq:currentpropagators}) for the relevant propagators, replacing $\hat{\tilde\chi}_P$ by the overdamped limit result obtained from (\ref{eq:chiPnum}) and replacing $\hat\chi({\mathbf q},t)_{\hat0\hat0}$ by the numerical solution of Eq.\ (\ref{eq:dchi00dt2}).

Every function used in this section has a corresponding self function, with the exception of the $Q$ vertex functions.  Every equation in this section has a `self' form that can be obtained by adding a subscript $s$ to the symbol for every function in the equation (with the exception of $Q$ functions) and deleting the $\hat Q_{11}^{c0}$ terms in Eqs.\ (\ref{eq:calQ}) and (\ref{eq:MO}).  Thus, for example, Eqs.\ (\ref{eq:MO}) and (\ref{eq:dchi00dt2}) become
\[\hat{\tilde M}_{s}^{O}({\mathbf q},z) \equiv\left[\hat Q_{11}^{c1}({\mathbf q})\hat{\chi}^{ED}_{sP}({\mathbf q})\hat{\tilde M}_{s\,red}({\mathbf q},z)\hat Q_{11}^{c1}({\mathbf q})\right]_{\hat0\hat0}
\]
\begin{eqnarray*}
\lefteqn{\frac{\partial \hat\chi_s(q\hat{\mathbf k},t)_{\hat0\hat0}}{\partial t} =
\delta(t)+\hat{\mathcal Q}_{s11}(q\hat{\mathbf k})\hat\chi_s(q\hat{\mathbf k},t)_{\hat0\hat0}}
\\*
&&+
\int_0^t dt^{\prime}\,\hat M^O_s(q\hat{\mathbf k},t-t^\prime)\hat\chi_{s}({\mathbf q},t^{\prime})_{\hat0\hat0}\quad\quad\quad\quad
\end{eqnarray*}
Similar adjustments are applicable to the equations for the current correlation functions to obtain those for the self current correlation functions.

In practice, the numerical work required to calculate the correlation functions of interest is simplified by the fact that the basic equations for the five correlation functions [see Eqs.\ (\ref{eq:phirhodef})-(\ref{eq:phijlsdef}), (\ref{eq:dchi00dt}), and (\ref{eq:currentpropagators})] require, as input, Fourier transforms of certain functions evaluated only for the wave vector ${\bf q}$ pointing in the $z$ direction.  As a result, the calculations can be performed for this special case only.  Moreover, the equations in this section have a structure such that the equations for different values of ${\bf q}$ are uncoupled and can be solved separately for each value of $|{\bf q}|$.  Finally, in the time domain (rather than the Laplace transform domain), the equations above become simple to deal with numerically. 

An example of such a calculation is discussed in the following paper.\cite{paper2}


\begin{thebibliography}{10}
\bibitem{NOTE2}
We use the terms `diagram' and `graph' interchangeably.
\end{thebibliography}
\end{document}